%% file: paper.tex
\documentclass[journal,10pt]{IEEEtran}

\usepackage{epsf}
\usepackage{graphicx}
\usepackage{epsfig,latexsym,amsmath,epsf,amssymb,amsfonts}
\usepackage{amssymb}
\usepackage{placeins}
\usepackage{subfiles}
\usepackage[linesnumbered,ruled,vlined]{algorithm2e}
\usepackage{silence}
\usepackage[hidelinks]{hyperref}
\hypersetup{pdfcreator={Me}}
\usepackage{graphicx}
\usepackage{epstopdf}
\usepackage{watermark}
\usepackage{fancyhdr}
\usepackage{makeidx}
\usepackage[toc,nonumberlist]{glossaries}
\usepackage[nottoc]{tocbibind}
\usepackage{amsmath}
\usepackage{amssymb}
\usepackage[capitalise]{cleveref}
\usepackage{amsthm}

\usepackage{mathtools}
\usepackage{breqn}
\usepackage{xargs}
\usepackage[obeyFinal,colorinlistoftodos,prependcaption]{todonotes}
\usepackage{breqn}
\usepackage{cuted}
\usepackage{subcaption}
\usepackage{changes}
\usepackage{mathtools}
\usepackage{verbatim}
\usepackage{physics}
\usepackage{stfloats,lipsum}
\usepackage{array}
\usepackage{cite}
\usepackage{float}

\newcolumntype{P}[1]{>{\centering\arraybackslash}p{#1}}

\usepackage{pifont}
\newcommand{\cxmark}{\ding{55}}

\usepackage{lipsum}

\hypersetup{pdfcreator={Me}}
\newcommand\blfootnote[1]{%
\begingroup
\renewcommand\thefootnote{}\footnote{#1}%
\addtocounter{footnote}{-1}%
\endgroup
}

\usepackage{mathtools}
\DeclarePairedDelimiter\ceil{\lceil}{\rceil}

\usepackage[utf8]{inputenc}
\usepackage[english]{babel}

\newtheorem{proposition}{Proposition}

\renewcommand\IEEEkeywordsname{Index Terms}

\usepackage[justification=justified, font = small]{caption}

\subfile{\main/abbrev}

\newcommand{\simwidth}{1\columnwidth}
\newcommand{\figref}[1]{Fig. \ref{#1}}

\begin{document}
\title{RIS-enabled Multi-user $M$-QAM Uplink NOMA Systems: Design, Analysis, and Optimization}
\author{\large Mahmoud AlaaEldin, \textit{Member}, \textit{IEEE}, Mohammad Al-Jarrah, \textit{Member}, \textit{IEEE}, Xidong Mu, \textit{Member}, \textit{IEEE}, Emad Alsusa, \textit{Senior Member}, \textit{IEEE}, Karim G. Seddik, \textit{Senior Member}, \textit{IEEE}, and Michail Matthaiou, \textit{Fellow}, \textit{IEEE}  \vspace{-0.25in} \\ 
\normalsize
}
\maketitle
\thispagestyle{plain}
\pagestyle{plain}

\begin{abstract}
Non-orthogonal multiple access (NOMA) is widely recognized for enhancing the energy and spectral efficiency through effective radio resource sharing. However, uplink NOMA systems face greater challenges than their downlink counterparts, as their bit error rate (BER) performance is hindered by an inherent error floor due to error propagation caused by imperfect successive interference cancellation (SIC). This paper investigates BER performance improvements enabled by reconfigurable intelligent surfaces (RISs) in multi-user uplink NOMA transmission. Specifically, we propose a novel RIS-assisted uplink NOMA design, where the RIS phase shifts are optimized to enhance the received signal amplitudes while mitigating the phase rotations induced by the channel. To achieve this, we first develop an accurate channel model for the effective user channels, which facilitates our BER analysis. We then introduce a channel alignment scheme for a two-user scenario, enabling efficient SIC-based detection and deriving closed-form BER expressions. We further extend the analysis to a generalized setup with an arbitrary number of users and modulation orders for quadrature amplitude modulation signaling. Using the derived BER expressions, we develop an optimized uplink NOMA power allocation (PA) scheme that minimizes the average BER while satisfying the user transmit power constraints. It will be shown that the proposed NOMA detection scheme, in conjunction with the optimized PA strategy, eliminate SIC error floors at the base station. The theoretical BER expressions are validated using simulations, which confirms the effectiveness of the proposed design in eliminating BER floors.
\end{abstract}


\blfootnote{Mahmoud AlaaEldin, Xidong Mu, and Michail Matthaiou are with the Centre for Wireless Innovation (CWI), Queen’s University Belfast, Belfast BT3 9DT, UK (e-mail: m.alaaeldin@qub.ac.uk; x.mu@qub.ac.uk; m.matthaiou@qub.ac.uk)

Mohammad Al-Jarrah and Emad Alsusa are with the Electrical and Electronic Engineering Department, University of Manchester, Manchester M13 9PL, UK (e-mail: mohammad.al-jarrah@manchester.ac.uk; e.alsusa@manchester.ac.uk).

Karim G. Seddik is with the Department of Electronics and Communications Engineering, American University in Cairo, Cairo, Egypt 11835 (e-mail: kseddik@aucegypt.edu). }

\vspace{-0.15in}



In recent years, there has been an unprecedented increase in the number of mobile terminals, \gls{iot} devices, and autonomous vehicles, placing a tremendous burden on the existing limited radio resources. Moreover, emerging technologies, such as holographic communications, virtual and augmented reality applications, video streaming, and online gaming, all of which require massive data rates, have further amplified the demand for efficient utilization of these resources. To address this challenge, \gls{noma} has been proposed as a revolutionary multiple-access technique that can in theory enhance the spectral efficiency. As a result, \gls{noma} has gained significant attention from both academia and industry and is regarded a key candidate for future sixth-generation (6G) networks and beyond \cite{9022993, 9154358, 7676258}. \gls{noma} is categorized into two main schemes: power-domain \gls{noma} and code-domain \gls{noma}. In the former, users are assigned different power levels based on their channel conditions to the \gls{bs}, while in the latter, each user is assigned a unique code as a signature. This paper focuses on power-domain \gls{noma}, where multiple devices share the same time-frequency resource block with different power levels, and \gls{sic} is employed at the receiver to effectively decode the signals \cite{hamad_surv}.

Likewise, \glspl{ris} have emerged in recent years as smart technology capable of improving the quality of wireless communication links \cite{mos1, yuanwei_surv, ris_beamforming, ris_iot_surv, Qingqing}. An \gls{ris} typically consists of a large number of low-cost, tunable passive elements that can adjust the amplitudes and phase shifts of incident waves, effectively reflecting or refracting them toward the receiver. The key advantage of \glspl{ris} lies in their ability to customize the wireless channel to support various critical features, such as signal enhancement and interference mitigation. Consequently, the integration of \glspl{ris} into diverse wireless systems has been extensively studied in recent literature. For example, in \cite{Qingqing2}, the authors developed alternation minimization algorithms to minimize the total transmit power by jointly optimizing the active beamforming at the \gls{bs} and the passive beamforming at the \gls{ris}. In \cite{Debbah}, the maximization of the \gls{ee} of \gls{ris}-assisted downlink systems was considered. Furthermore, a joint power and user association scheme for a multi-\gls{ris}-assisted multi-\gls{bs} system using millimeter waves was proposed in \cite{user_assoc}.

\subsection{Related work}
Since \gls{noma} is capable of approaching the boundary of the capacity region, multiple works in the literature have studied the integration of \gls{noma} in \gls{ris}-enabled systems as a win-win solution \cite{ris_noma_surv, ris_noma_interplay}. In the following, we provide a review of the literature focusing on the performance analysis of uplink \gls{ris}-\gls{noma} systems.

The authors of \cite{tahir_1} studied the outage performance of a simplified \gls{ris} enabled uplink two-user \gls{noma} system. Specifically, they derived approximate analytical expressions for the received powers of the \gls{noma} users as Gamma \glspl{rv} using the method of moment matching, which resulted in tractable expressions of the outage probability with \gls{sic}. In addition, the study in \cite{tahir_2} analyzed the outage performance of a \gls{ris}-assisted two-user uplink \gls{noma} system, which divided the \gls{ris} elements into two groups to improve each user's signal. By approximating the received powers as Gamma \glspl{rv}, they derived outage probability expressions for uplink \gls{noma} signaling using \gls{sic}. The authors of \cite{yuanwei_1} investigated downlink and uplink \gls{ris}-aided \gls{noma} and \gls{oma} systems, where an \gls{ris} is applied to enhance the coverage for a cell-edge user communicating with the \gls{bs}. Closed-form expressions for the outage probability and ergodic rate under Nakagami-$m$ fading have been derived. Using asymptotic approximations in the high \gls{snr} regime, it has been revealed that there is a diversity order that depends on the number of \gls{ris} elements and fading parameters, while the high \gls{snr} slope is unaffected. In \cite{ser_star_ris}, the authors studied the \gls{ser} performance of \gls{star-ris} aided uplink \gls{noma}, where the phase shift and amplitude parameters were adjusted based on distance-maximizing constellation scaling and rotation to enhance the \gls{ser} performance of the system. However, closed-form \gls{ser} expressions were not derived in \cite{ser_star_ris} as the expressions are given by $K$-fold integrals. Finally, the authors of \cite{arman} investigated the ergodic sum-rate of the \gls{star-ris} assisted uplink \gls{noma} under channel estimation errors and hardware impairments, where upper bounds were derived for perfect and imperfect \gls{sic}-based decoding.

\subsection{Motivations and Contributions}

\begin{table}[t]
\begin{center}
\renewcommand{\arraystretch}{1.15}
\begin{tabular}{c c c c c c c c} 
\hline
  & Our work & \cite{tahir_1} & \cite{tahir_2} & \cite{yuanwei_1} & \cite{ser_star_ris} & \cite{arman}   \\ [0.5ex] 
 \hline\hline
 \gls{ris}-\gls{noma} & \checkmark & \checkmark & \checkmark & \checkmark & \checkmark & \checkmark \\
 \hline
 Uplink & \checkmark & \checkmark & \checkmark & \checkmark & \checkmark & \checkmark  \\
 \hline
 BER analysis & \checkmark & \cxmark & \cxmark & \cxmark & \cxmark & \cxmark   \\
 \hline
 Arbitrary \# of users & \checkmark & \cxmark & \cxmark & \cxmark & \cxmark & \cxmark  \\
 \hline
 Arbitrary Mod. orders & \checkmark & \cxmark & \cxmark & \cxmark & \cxmark & \cxmark  \\
 \hline
 BER based uplink PA & \checkmark & \cxmark & \cxmark & \cxmark & \cxmark & \cxmark  \\ 
 \hline
\end{tabular}
\caption{Distinguishing the work in this paper from other existing works in the literature}
\label{table_1}
\vspace{-0.2 in}
\end{center}
\end{table}

From the literature review, it is evident that most of the existing works have focused on the analysis of the achievable rate and outage capacity of \gls{ris}-enabled uplink \gls{noma} systems. To the best of the authors' knowledge, the investigation of reliability metrics, such as \gls{ber}, has not been considered in the literature. Studying the \gls{ber} performance analysis is indispensable for characterizing the reliability of communication systems, as it helps to predict the error rate of the received data, which has a direct impact on the effective throughput of communication links. Therefore, to bridge this clear gap in the literature, this paper provides a comprehensive study of the \gls{ber} performance limits of uplink \gls{ris}-\gls{noma} systems under imperfect \gls{sic}. Similar to existing literature \cite{lina_iot, tahir_1, tahir_2, arman}, the \gls{ris} is partitioned into $K$ partitions, where $K$ is the number of users, and each partition is assigned to a certain user. Then, the phase shifts of each \gls{ris} partition are adjusted to maximize the effective channel of the corresponding user. This partitioning technique allows us to statistically model the effective channels of the users which is required to derive the average \gls{ber} analytical expressions.

The key contributions of this work are summarized as follows:

\begin{itemize}

\item A detailed and accurate statistical modeling of the effective channels of the users is introduced based on dividing the \gls{ris} panel into $K$ partitions and assigning each partition to a certain user.

\item A novel channel alignment scheme is then devised to align the effective channels of the users on the real axis to eliminate the relative phase shifts between the effective channels of the users. This allows for the construction of a simple and tractable SIC-based detection of the users' symbols at the \gls{bs}. The proposed channel alignment facilitates the \gls{ber} performance analysis of the two-user uplink \gls{ris}-\gls{noma} scenario.

\item An extension of the \gls{ber} analysis is conducted for the case of arbitrary number of users and \gls{qam} modulation orders, providing valuable insights on the achievable performance of the \gls{ris}-\gls{noma} system. This is done by devising an algorithm that computes the coefficients of the \gls{ber} expressions for any system with arbitrary parameters.

\item A \gls{pa} scheme that utilizes the derived \gls{ber} expressions is devised to eliminate the \gls{ber} floors, by minimizing the average \gls{ber} of all users subject to individual uplink transmit power constraints for the users.

\item Simulation results verify the accuracy and effectiveness of the proposed closed-form generic analytical \gls{ber} expressions. The results also reveal the efficacy of the proposed \gls{pa} scheme, showing that it completely eliminates the \gls{ber} floors. Moreover, we compare \gls{ris}-\gls{noma} systems with the equivalent \gls{ris}-\gls{oma} counterparts. The results confirm the superiority of \gls{ris}-\gls{noma} over \gls{ris}-\gls{oma} across all scenarios and almost all \gls{snr} values when both systems maintain the same channel estimation overhead for a fair comparison. However, if the additional channel estimation, computational demands, and hardware complexities of \gls{ris}-\gls{oma} are ignored, it may slightly outperform \gls{ris}-\gls{noma} in cases with low modulation orders and a limited number of reflectors. Nonetheless, this comparison would not be fair.

\end{itemize}

Finally, Table \ref{table_1} explicitly contrasts our contributions to the existing works on \gls{ris}-\gls{noma} in the literature.

\subsection{Organization and Notations}
The rest of the paper is organized as follows: In Sec. \ref{sysmod}, the \gls{ris}-enabled uplink \gls{noma} system model is presented. The statistical modeling and channel alignment of the \gls{ris} uplink channels is presented in Sec. \ref{stat_model}. In Sec. \ref{2-user}, we present the \gls{ber} analysis of the two-user uplink \gls{ris}-\gls{noma} system, whereas the analysis is extended in Sec. \ref{general} for the generalized system with an arbitrary number of \gls{noma} users and arbitrary modulation orders. An optimized uplink \gls{pa} scheme for the uplink \gls{ris}-\gls{noma} system is discussed in Sec. \ref{opt}. The simulation results and discussion are given in Sec. \ref{Sim}, while the conclusions of this work are presented in Sec. \ref{Conc}.

\textit{Notations:}
The bold lowercase letters are used to define vectors, while the bold uppercase letters are used to define matrices. The $|\cdot|$, $\mathfrak{Re}(\cdot)$, $\mathfrak{Im}(\cdot)$, $\arg(\cdot)$, and $(\cdot)^*$ are the absolute, real, imaginary, angle, and conjugate of a complex number, while $j^\prime = \sqrt{-1}$ is the imaginary number.




\section{System Model}  \label{sysmod}

\begin{figure}
\centering
\includegraphics[width=0.9\simwidth]{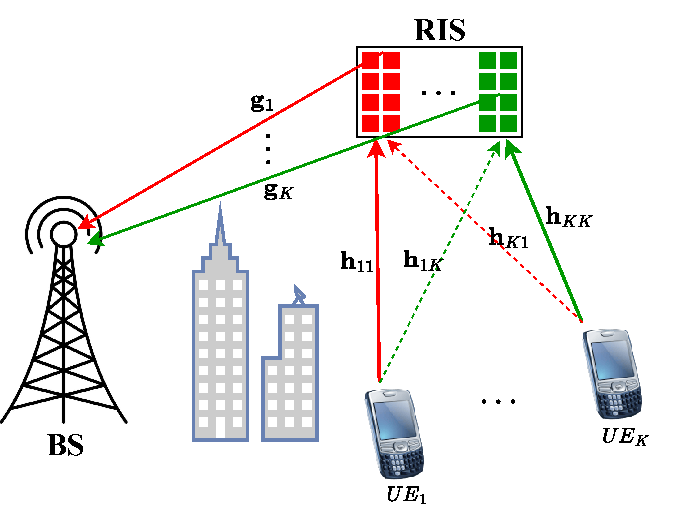}
\caption{RIS-enabled uplink NOMA system model}
\vspace{-0.1in}
\label{sys_model}
\end{figure}

As shown in \figref{sys_model}, we consider an uplink \gls{ris}-enabled \gls{noma} system that contains a single antenna \gls{bs}, an \gls{ris} with $L$ reflectors, and $K$ single antenna users, where the $i$th user is denoted by $U_i$. In our proposed design, we assume that the \gls{ris} is divided into separate $K$ partitions, denoted as $\mathcal{P}_1, \mathcal{P}_2, \cdots, \mathcal{P}_K$, where partition $\mathcal{P}_i$ is assigned to $U_i$ and contains $L_i$ reflection elements of the \gls{ris}, i.e., $\sum_{k=1}^KL_i=L$. The \gls{ris} phase shifts of each partition are adjusted to align the channels of the corresponding user so that the received signal components of each user are coherently added at the destination. As shown in \figref{sys_model}, the channel vector from $U_i$ to its associated \gls{ris} partition, $\mathcal{P}_i$, is denoted as $\mathbf{h}_{ii} \in \mathbb{C}^{L_i {\times} 1}$. On the other hand, the cross channel vector from $U_i$ to the partition $\mathcal{P}_j$ is denoted as $\mathbf{h}_{ij} \in \mathbb{C}^{L_j {\times} 1}$. Moreover, the channel vector from the $i$th portion, $\mathcal{P}_i$, to the \gls{bs} is denoted as $\mathbf{g}_i \in \mathbb{C}^{L_i {\times} 1}$. Therefore, the received superimposed \gls{noma} signal, $y$, at the \gls{bs} can be expressed as
\begin{equation} \label{rx_sig}
y = \sum_{i=1}^K \sum_{j=1}^K \sqrt{ \frac{P_i \eta_i}{\beta_i} } \mathbf{h}_{ij}^T \mathbf{\Theta}_j \mathbf{g}_j x_i + n,
\end{equation}
where $P_i$ denotes the transmit power of $U_i$, $\eta_i$ represents the overall path loss of the cascaded channel, i.e., user-\gls{ris}-\gls{bs} link, and $\mathbf{\Theta}_j$ is a diagonal matrix whose diagonal elements are the $\mathcal{P}_j$ reflection coefficients, i.e., $\mathbf{\Theta}_j = \mathrm{diag} \{e^{j^\prime \theta_{j,1}}, e^{j^\prime \theta_{j,2}}, \dots, e^{j^\prime \theta_{j,L_j}} \}$, with $\theta_{j,l}$ representing the phase shift introduced by the $l$th reflecting element of $\mathcal{P}_j$. The term $x_i \in \mathbb{C}$ denotes the transmitted modulation symbol of $U_i$, drawn from a square \gls{qam} alphabet, $\mathcal{X}_i$, where the cardinality of $\mathcal{X}_i$ is the \gls{qam} modulation order of $U_i$. Since the data symbols, $x_i$, are drawn from square \gls{qam} constellations, both the real and imaginary components of $x_i$ can take values from the set $\{ \pm 1, \pm 3, \dots, \pm \sqrt{M_i}-1 \}$, where $M_i$ is the modulation order of $U_i$. The scaling factor, $\beta_i\triangleq\frac {2}{3} \left( {M_i - 1} \right)$, is used to normalize $x_i$ to unity. The parameter, $\eta_i= \eta_{U_i,I} \eta_{I,B}$, in (\ref{rx_sig}) is the overall path loss of the $U_i$-\gls{ris}-\gls{bs} link, where $\eta_{U_i,I} = d_{U_i,I}^{-\psi}$, $\eta_{I,B} = d_{I,B}^{-\psi}$; $d_{U_i,I}$ is the distance between the $i$th user and \gls{ris}, $d_{I,B}$ is the \gls{bs}-\gls{ris} distance, where $\psi$ is the path loss exponent of the wireless links. The additive white Gaussian noise at \gls{bs} is denoted by $n$, and it is modeled as a complex Gaussian \gls{rv} with zero mean and variance of $\sigma_n^2$ \textit{per complex dimension}, i.e., $n \sim \mathcal{CN}(0, 2\sigma_n^2)$.

The channel vectors $\mathbf{g}_j$ and $\mathbf{h}_{ij}, \forall i, j$ are assumed to be Rayleigh fading channels. Therefore, the elements of $\mathbf{h}_{ij}$ and $\mathbf{g}_j$ are assumed as \gls{iid} complex normal \glspl{rv} with zero mean and unit variance, i.e., $\mathbf{h}_{ij}, \mathbf{g}_j \sim \mathcal{CN}(\mathbf{0}, \mathbf{I})$. The \gls{ris} is assumed to have full knowledge of the phases of $\mathbf{h}_{ii}$ and $\mathbf{g}_i, \forall i$, so that the phase shifts of the \gls{ris} reflection elements can be adjusted to reflect incident signals with concentrated beams towards the \gls{bs}. It should be noted that there is no need for the estimation of the channels of all \gls{ris} elements to the users, i.e., knowledge of the cross channel vectors, $\mathbf{h}_{i,j}, \forall i \neq j$, is not required. Since the $i$th user, $U_i$, is assigned a separate partition of the \gls{ris}, $\mathcal{P}_i$, the phase shifts of $\mathcal{P}_i$ are adjusted to maximize the reflected signal from $\mathcal{P}_i$ at the \gls{bs}. However, $U_i$ signal reflections from other \gls{ris} partitions, $\mathcal{P}_j, \forall j \neq i$, are considered random signal reflections, since the phase shifts of these partitions are not adjusted to the $U_i$ channels. Therefore, two signal components of $U_i$ are received at the \gls{bs}, namely, the optimized component and the random component. To maximize the reflected signal component of $\mathcal{P}_i$ received at the \gls{bs}, the adjustable phase shifts of $\mathcal{P}_i$ are set as $\theta_{i,l} = -(\arg(h_{ii,l}) + \arg(g_{i,l}))$, $\forall l = 1, \dots, L_i$, where $\theta_{i,l}$ is the phase shift of the $l$th reflection element of $\mathcal{P}_i$, and $h_{ii,l}$ and $g_{i,l}$ are the $l$th elements of $\mathbf{h}_{ii}$ and $\mathbf{g}_i$, respectively. As a result, the total received superimposed signal at the \gls{bs} can be written as
\begin{equation}  \label{rx_sig_modified}
y = \sum_{i=1}^K \left( \gamma_{ii} + \sum_{j=1, j \neq i}^K  \gamma_{ij} \right) x_i + n,
\end{equation}
where $\gamma_{ii}$ is the optimized component of the \gls{ris} effective channel of the $i$th user, reflected from its allocated partition, $\mathcal{P}_i$, which can be expressed as
\begin{equation}  \label{opt_comp}
\gamma_{ii} = \sqrt{ \frac{P_i \eta_i}{\beta_i} } \sum_{l=1}^{L_i} |h_{ii,l}| |g_{i,l}|.
\end{equation}
The terms, $\gamma_{ij}, \forall j \neq i$, are the random components of the \gls{ris} effective channel of user $i$, reflected from the other users' partitions, $\mathcal{P}_j, \forall j \neq i$, which can be expressed as
\begin{equation}  \label{rand_comp}
\gamma_{ij} = \sqrt{ \frac{P_i \eta_i}{\beta_i} } \sum_{l=1}^{L_j}  h_{ij,l} \ e^{j^\prime \theta_{j,l}} \ g_{j,l},
\end{equation}
where $h_{ij,l}$ is the $l$th element of $\mathbf{h}_{ij}$, and $g_{j,l}$ is the $l$th element of $\mathbf{g}_j$.

In the next section, we provide an accurate statistical modeling for the \gls{ris} effective channels and a channel alignment scheme to align the effective channels of the users. This will be used to derive closed-form average \gls{ber} expressions in the subsequent sections.


\section{Statistical modeling and alignment of the RIS effective channels of the users}  \label{stat_model}

In this section, accurate statistical modeling is derived for both the optimized component, $\gamma_{ii}$, and the random components, $\gamma_{ij}$, of the \gls{ris} effective channel of the $i$th user. Moreover, a channel alignment scheme is devised to align the effective channels of the users to the zero axis, which later facilitates simple detection and the \gls{ber} derivations. The statistical models of the channel components derived in this section will be used in Sec. \ref{2-user} and Sec. \ref{general} for the calculation of the unconditional average \gls{ber} of the users at \gls{bs}.

\subsection{Statistical modeling of the effective channels of the users}
The optimized component, $\gamma_{ii}$, is a summation of cascaded \gls{iid} Rayleigh \glspl{rv}, as in (\ref{opt_comp}), which can be modeled as a Gamma \gls{rv} using moment matching according to the casual form of the \gls{clt} \cite{papoulis}. Since both $h_{ii,l}$ and $g_{i,l}$ in (\ref{opt_comp}) are independent complex normal \glspl{rv} with unit variances, the mean value of $\gamma_{ii}$ can be derived using the mean formula of Rayleigh \gls{rv} \cite{papoulis} as
\begin{align}
\mathbb{E}(\gamma_{ii})  = \sqrt{ \frac{P_i \eta_i}{\beta_i} } L_i \mathbb {E}(|h_{ii,l}| |g_{i,l}|)   = \sqrt{ \frac{P_i \eta_i}{\beta_i} } \frac{\pi}{4} L_i.
\end{align}
Moreover, since the variance of the product of two Rayleigh \glspl{rv}, $|h_{ii,l}| |g_{i,l}|$, when both $|h_{ii,l}|$ and $|g_{i,l}|$ are $\mathcal{CN}(0,1)$, is $\frac{16 - \pi^2}{16}$ \cite{alaaeldin1}, the variance of $\gamma_{ii}$ can be calculated as
\begin{equation}
\mathrm{Var}(\gamma_{ii}) = \frac{P_i \eta_i}{\beta_i} \frac{16 - \pi^2}{16} L_i.
\end{equation}
Let us define the scale and shape parameters of the equivalent Gamma \gls{rv} by $\zeta$ and $N$, respectively. Then, $\gamma_{ii}$ has a mean value of $N \zeta$ and a variance of $N \zeta^2$ \cite{papoulis}. Therefore, by equating the mean and variance of $\gamma_{ii}$ to the mean and variance of the Gamma \gls{rv}, both the scale and shape parameters of the Gamma \gls{rv} can be derived as
\begin{align}  \label{zeta_and_N}
 \quad \zeta_i &= \sqrt{ \frac{P_i \eta_i}{\beta_i} } \frac{16 - \pi^2}{4 \pi}, \quad N_i = L_i \frac{\pi^2}{16 - \pi^2}.
\end{align}
Finally, the \gls{cf} of $\gamma_{ii}$ can be given as
\begin{equation} \label{gamma_approx}
\Phi_{\gamma_{ii}}(z) = \left( 1 - j^\prime \zeta_i z \right) ^{-N_i}.
\end{equation}

On the other hand, the random components of the effective uplink channels, $\gamma_{ij}$, have different statistical distributions, which we derive in the following. Without loss of generality, by setting $P_i$, $\eta_i$ and $\beta_i$ to unity, the normalized random components of the uplink channels, $\widehat{\gamma}_{ij}$, can be expressed as
\begin{equation}  \label{normalized_rand_comp}
\widehat{\gamma}_{ij}  =  \sum_{l=1}^{L_j}  h_{ij,l} \ e^{j^\prime \theta_{j,l}} \ g_{j,l},  \quad  \forall i,j, \ i \neq j.
\end{equation}
Clearly, $\widehat{\gamma}_{ij}$ is an $L_j$-sum of the product of independent complex \gls{rv}s, which makes $\widehat{\gamma}_{ij}$ another complex \gls{rv} having real and imaginary components. The following proposition gives the statistical distribution of both the real and imaginary components of $\widehat{\gamma}_{ij}$.
\begin{proposition}  \label{theo_1}
The real and imaginary parts of $\widehat{\gamma}_{ij}$ are identically distributed \glspl{rv} where each of them can be modeled as a difference between two \gls{iid} Erlang distributions with shape and scale parameters of $L_j$ and $0.5$, respectively. Therefore, the \gls{cf} of $\mathfrak{Re}(\widehat{\gamma}_{ij})$ can be given as
\[ \Phi_{\mathfrak{Re}(\widehat{\gamma}_{ij})}(z) = \left( 1 + 0.25 z^2 \right)^{-L_j} .\]
\end{proposition}

\renewcommand\qedsymbol{$\blacksquare$}

\begin{proof}
See Appendix \ref{appx_a}.
\end{proof}

\begin{figure}
\centering
\includegraphics[width=0.9\simwidth]{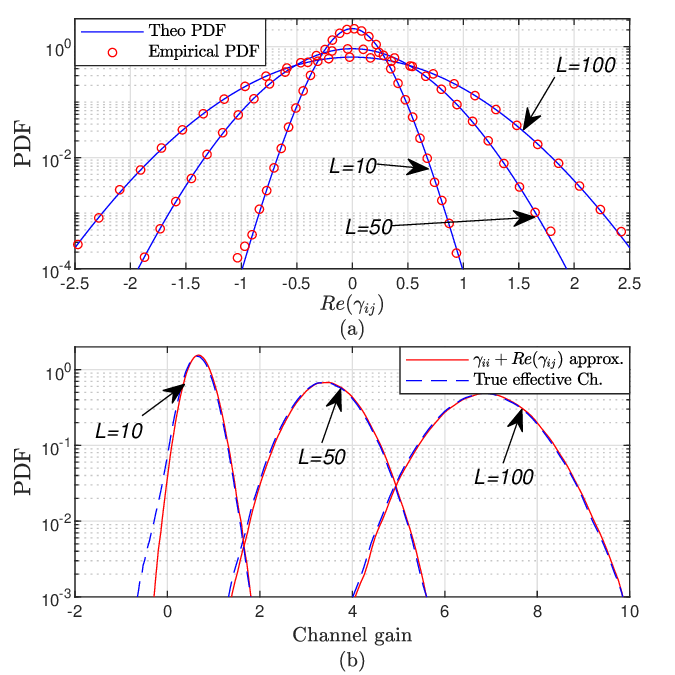}
\caption{Comparison between the empirical and theoretical channels' PDFs of a two-user system under different values of $L$, $d_{U_1,I} = 20$, $d_{U_2,I} = 50$, $d_{I,B} = 30$, $\psi = 2.2$, and $P_1{=}P_2{=}40$ dB: (a) PDF of $\mathfrak{Re} (\gamma_{ij})$, (b) PDF of the effective channel of $U_i$ after channel alignment.}
\vspace{-0.1in}
\label{pdf_fig}
\end{figure}

Since $\gamma_{ij}$ is just a scaled version of $\widehat{\gamma}_{ij}$ with a scale factor of $\sqrt{\frac{P_i \eta_i}{\beta_i}}$, then the \gls{cf} of $\mathfrak{Re}(\gamma_{ij})$ can be directly found by replacing every $z$ with $\sqrt{\frac{P_i \eta_i}{\beta_i}} z$ in $\Phi_{\mathfrak{Re}(\widehat{\gamma}_{ij})}(z)$ as
\begin{equation}  \label{diff_erlang}
\Phi_{\mathfrak{Re}(\gamma_{ij})}(z) = \Phi_{\mathfrak{Re}(\widehat{\gamma}_{ij})} \left( \sqrt{\frac{P_i \eta_i}{\beta_i}} z \right)  {=}  \left( 1 + \frac{P_i \eta_i}{4 \beta_i} z^2 \right)^{-L_j}.
\end{equation}
Figure \figref{pdf_fig}(a) confirms that the derived theoretical \gls{pdf} of $\mathfrak{Re}(\gamma_{ij})$ perfectly matches the empirical \gls{pdf} using simulations with different values of $L$.


\subsection{Channel alignment scheme for the users' effective channels}  \label{ch_align}
In this subsection, we present an approach to align the overall effective channel of each user to the real axis by canceling the relatively small imaginary random components of the channel. This alignment scheme facilitates the efficient and simple detection of the uplink transmit data symbols of the users since it allows for simpler decision regions and decision thresholds. Moreover, obtaining closed-form expressions for \gls{ber} becomes feasible using the proposed channel alignment scheme, since the received signal constellation of the superimposed \gls{noma} symbols at the \gls{bs} becomes a \gls{qam}-like constellation, allowing tractable \gls{ber} analysis.

To align the effective channels of the users in (\ref{rx_sig_modified}), let us define the angles, $\alpha_j, \forall j \in \{1, \dots, K\}$, as two control angles at the \gls{ris}. Specifically, $\alpha_j$ is set to control the \gls{ris} partition, $\mathcal{P}_j$, by rotating all its elements by the same value of $\alpha_j$. Therefore, the received superimposed \gls{noma} signal at the \gls{bs} can be expressed as
\begin{equation}
y = \sum_{i=1}^K \sum_{j=1}^K \gamma_{ij} e^{j^\prime \alpha_j} x_i + n.
\end{equation}
The angles $\alpha_j, \forall j$ are set so that they cancel the imaginary components of the overall effective channels of all the users, $h_i^{eff}, \forall i {\in} \{1, \dots, K\}$. Hence, $\alpha_j, \forall j$ are calculated by simultaneously solving the following equations
\begin{equation}  \label{align}
\mathfrak{Im} \left( \sum_{j=1}^K \gamma_{ij} e^{j^\prime \alpha_j} \right) = 0, \quad \forall i \in \{1, 2, \dots, K\},
\end{equation}
where the system of equations in (\ref{align}) consists of $K$ nonlinear equations in $K$ unknown angles, $\alpha_j, \forall j$, which can be numerically solved using a numerical technique such as Newton's method.

Since the optimized components, $\gamma_{ii}, \forall i$, have much higher magnitude than the random components, $\gamma_{ij}, \forall i \neq j$, the solution for (\ref{align}) would normally have small values for $\alpha_j, \forall j$. In other words, the overall effective channels of the users can be aligned by adding a slight phase rotation to each \gls{ris} partition, which allows the large optimized channel components, $\gamma_{ii}, \forall i$, to dominate the small random components, $\gamma_{ij}, \forall i \neq j$. Thus, the overall effective channels of the users, $h_i^{eff}, \forall i$, become purely real after applying the control angles, $\alpha_j$, and they can be written as
\begin{equation} \label{real_approx}
h_i^{eff} {=} \sum_{j=1}^K \gamma_{ij} e^{j^\prime \alpha_j} \approx  \gamma_{ii} + \sum_{ \substack{j=1 \\ j \neq i} }^K \Re(\gamma_{ij}), \quad \forall i \in \{1, \dots, K\},
\end{equation}
where $\approx$ means approximately equal. This approximation is validated using Monte-Carlo simulations in \figref{pdf_fig}(b), where we compare the true empirical \gls{pdf} of the effective channel to the approximate statistical model in (\ref{real_approx}). Therefore, the \gls{cf} of $h_i^{eff}, \forall i$, can be given using (\ref{gamma_approx}) and (\ref{diff_erlang}) as
\begin{equation}
\Phi_{h_i^{eff}}(z)  \approx  \Phi_{\gamma_{ii}}(z) \prod_{j=1, j \neq i}^K  \left( 1 + \frac{P_i \eta_i}{4 \beta_i} z^2 \right)^{-L_j}.
\end{equation}
The final received signal at the \gls{bs} can then be expressed as
\begin{equation}  \label{rx_sig_mult_user}
y = \sum_{i=1}^K h_i^{eff} x_i + n,
\end{equation}
where $h_i^{eff}, \forall i$ are purely real \glspl{rv} with zero phase rotations. In the next subsection, we show how the proposed channel alignment scheme facilitates the detection process and the \gls{ber} analysis.


\section{BER analysis of the two-user uplink RIS-NOMA system}  \label{2-user}

In this section, we provide a comprehensive \gls{ber} analysis for the two-user \gls{ris}-\gls{noma} scenario, assuming $16$-\gls{qam} and \gls{qpsk} modulation for $U_1$ and $U_2$, respectively. Gray coding is employed to map binary bits to \gls{qpsk} symbols, with the mapping as follows: $00 \rightarrow s_0$, $01 \rightarrow s_1$, $10 \rightarrow s_2$, and $11 \rightarrow s_3$. Each \gls{qpsk} symbol's first and second bits are denoted as $b_{n1}$ and $b_{n2}$ respectively, where $n \in \{1, 2\}$ represents the $n$th user.

The symbols, $x_1$ and $x_2$, are jointly decoded from (\ref{rx_sig_mult_user}) based on the construction of the received superimposed \gls{noma} symbol constellation at the \gls{bs}. Given that the receiver noise, $n$, is circularly symmetric Gaussian, optimal detection simplifies to a minimum distance decoder based on the effective channel values, $h_i^{eff}, \forall i$. With zero phase rotations of the effective channels, the decision regions for the superimposed \gls{noma} symbols are simple rectangular shapes, akin to standard \gls{qam} detection. This \gls{qam}-like constellation structure facilitates simple and tractable data decoding through simple thresholds and decision regions. Therefore, the decision thresholds can be calculated given $h_1^{eff}$ and $h_2^{eff}$, as we show below. It should be noted that constructing simple and tractable decision regions facilitates the derivation of closed-form \gls{ber} analytical expressions, which will be used later in optimizing the uplink \gls{pa} to mitigate the ambiguity of the superimposed constellation of uplink \gls{noma} and its associated \gls{ber} floors.

\begin{figure*}
\centering
\includegraphics[width=2\columnwidth]{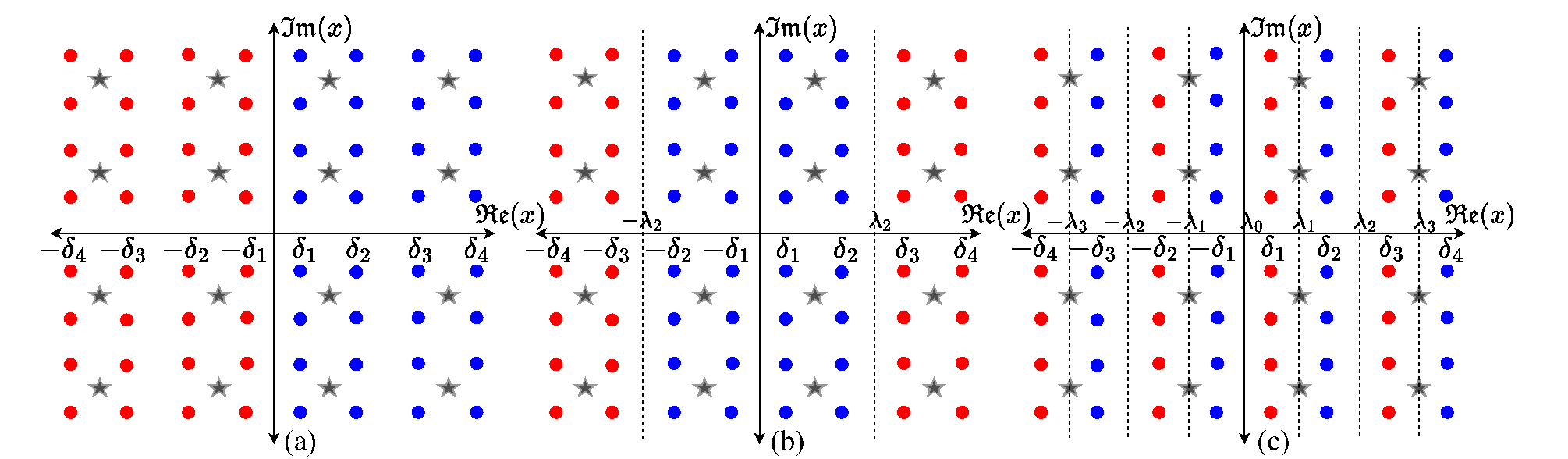}
\caption{Combined uplink NOMA constellations for the $2$-user $16$-$4$ QAM system: (a) $b_{11}$, (b) $b_{12}$, (c) $b_{21}$.}
\vspace{-0.1in}
\label{16_4_sys}
\end{figure*}

The constellation of the $16$-$4$-\gls{qam} \gls{noma} symbols of the two-user system is shown in \figref{16_4_sys}. The combined constellation diagram of the \gls{noma} symbol in this case has $64$ constellation points containing an information amount of $6$ bits per symbol, that is, four bits for $U_1$ and two bits for $U_2$. Figure \ref{16_4_sys}(a) divides the constellation points into two sets according to the value of $b_{11}$ equal to $1$ or $0$, that is, the constellation points that have $b_{11} = 0$ are red colored, while the remaining points having $b_{11} = 1$ are in blue. Figure \ref{16_4_sys}(b) divides the constellation diagram in the same way as \figref{16_4_sys}(a) but according to $b_{12}$. For $b_{13}$ and $b_{14}$, the constellation points are divided exactly in the same way as $b_{11}$ and $b_{12}$, respectively, but horizontally not vertically. On the other hand, \figref{16_4_sys}(c) shows the division of the constellation points according to the value of $b_{21}$ being $1$ or $0$. For the second bit of $U_2$, $b_{22}$, the constellation diagram is divided in the same way as in \figref{16_4_sys}(c) but horizontally not vertically.

Since \figref{16_4_sys} clearly divides the constellation into decision regions according to each bit, we will use it to derive the average \gls{ber} expressions for both $U_1$ and $U_2$ using a set of decision thresholds denoted by $\lambda_i$ which can be given as
\begin{equation}
\lambda_0 = 0, \quad \lambda_i = \frac{\delta_i + \delta_{i+1}}{2}, \forall i \in \{1,2,3\},
\end{equation}
where the values of the $\delta$'s are given as
\begin{align}
& \delta_1 = h_1^{eff} - h_2^{eff}, \quad \delta_2 = h_1^{eff} + h_2^{eff},  \nonumber  \\
& \delta_3 = 3 h_1^{eff} - h_2^{eff}, \quad \delta_4 = 3 h_1^{eff} + h_2^{eff}.
\end{align}
In the following \gls{ber} calculations, we assume that $\delta_4 > \lambda_3 > \delta_3 > \lambda_2 > \delta_2 > \lambda_1 > \delta_1>0$. Although these values depend on the values of the effective channels, $h_i^{eff}$, which are \gls{rv}s by nature, we derive all the \gls{ber} expressions assuming this desired order. The rationale of this choice is driven by the fact that this order can be achieved with optimized uplink \gls{pa} based on minimizing the obtained \gls{ber} expressions. This uplink \gls{pa} will ensure that the correct order is satisfied by reducing the channel ambiguity among the users, as it heavily penalizes the average \gls{ber} cost function.

\subsection{Analysis of \texorpdfstring{$U_1$}{Lg}}
To calculate the average \gls{ber} of $U_1$, we need to calculate the average probability of detection error of its individual bits, $b_{1i}, \forall i \in \{1, \dots, 4\}$, and take the average over these probabilities. Let us assume that the columns of the constellation diagrams in \figref{16_4_sys} are numbered from left to right as $C_1, \dots, C_8.$ Starting with $b_{11}$, a detection error can occur if any of the blue constellation points in \figref{16_4_sys}(a) moves to the negative side of the real axis or if any of the red points moves to the positive side. For example, assuming that the true received constellation point belongs to the fifth column, $C_5$, and given $h_i^{eff}$, the conditional error probability of $b_{11}$ can be expressed as
\begin{equation}
P_{b_{11}| h_i^{eff}, C_5} = \mathrm{Pr} \left(\mathfrak{Re}(n) < -\delta_1 \right) = Q \left( \frac{\delta_1}{\sigma_n} \right),
\end{equation}
where $Q(x) \triangleq \frac{1}{\sqrt{2 \pi}} \int_x^{\infty} e^{-u^2/2} du$. Similarly, the conditional error probability of $b_{11}$ can be calculated, given that the received constellation point belongs to any of the other columns, $C_m, \forall m$. Hence, the average bit error probability of $b_{11}$, averaged over all $C_m, \forall m {\in} \{1, \dots, 8\}$, can be calculated as
\begin{equation}  \label{b11}
P_{b_{11}| h_i^{eff}} = \frac{1}{4} \sum_{m=1}^4 Q \left( \frac{\delta_m}{\sigma_n} \right).
\end{equation}

On the other hand, we can notice from \figref{16_4_sys}(b) that a detection error can occur in detecting the second bit of $U_1$, $b_{12}$, if any of the blue points exceeds the decision region bounds, that is, from $-\lambda_2$ to $\lambda_2$ on the real axis, or if any of the red points enters that region. Hence, to calculate the detection error probability of $b_{12}$, we calculate the conditional error probability given that the received constellation point belongs to each of the columns, and then we take the average over all possible columns.

Given that the true received constellation point belongs to $C_1$, the conditional error probability of detecting $b_{12}$ can be calculated as
\begin{align}
P_{b_{12}| h_i^{eff}, C_1} & = \mathrm{Pr} \left(- \lambda_2 + \delta_4 < \mathfrak{Re}(n) < \lambda_2 + \delta_4 \right)  \nonumber  \\
& = Q \left( \frac{- \lambda_2 + \delta_4}{\sigma_n} \right) - Q \left( \frac{\lambda_2 + \delta_4}{\sigma_n} \right).
\end{align}
Similarly, if the received constellation point belongs to $C_m$, the conditional error probability of detecting $b_{12}$ can be calculated as
\begin{align}
& P_{b_{12}| h_i^{eff}, C_m} = \mathrm{Pr} \Big( \mathfrak{Re}(n) > (-1)^{\ceil*{\frac m 2}}\lambda_2 + \delta_{5 - m}, \nonumber \\
& \qquad \mathfrak{Re}(n) < (-1)^{\ceil*{\frac m 2} + 1}\lambda_2 + \delta_{5 - m} \Big)  \nonumber   \\
& \qquad = (-1)^{\ceil*{\frac m 2}} Q \left( \frac{\lambda_2 + (-1)^{\ceil*{\frac m 2} + 1} \delta_{5-m}}{\sigma_n} \right)  \nonumber  \\
& \qquad + Q \left( \frac{(-1)^{\ceil*{\frac m 2}}\lambda_2 + \delta_{5-m}}{\sigma_n} \right), \quad  \forall m \in \{1,2,3,4\}.
\end{align}
Due to symmetry in \figref{16_4_sys}(b), it is clear that the cases of $C_5$, $C_6$, $C_7$ and $C_8$ are similar to the cases of $C_4$, $C_3$, $C_2$ and $C_1$, respectively. Therefore, the average detection error probability of $b_{12}$ can be obtained by averaging over the first four columns as
\begin{equation}  \label{b12}
P_{b_{12}| h_i^{eff}} = \frac{1}{4} \sum_{m=1}^4  P_{b_{12}| h_i^{eff}, C_m}.
\end{equation}
Finally, since $b_{13}$ and $b_{14}$ are similar to $b_{11}$ and $b_{12}$, the conditional average \gls{ber} of $U_1$ given the values of the effective channels, $h_i^{eff}$, can be calculated by taking the average of the error probabilities of $b_{11}$ and $b_{12}$ as
\begin{equation}  \label{ber_U1_2users}
P_{U_1 | h_i^{eff}} = 0.5 (P_{b_{11}| h_i^{eff}} + P_{b_{12}| h_i^{eff}}),
\end{equation}
where $P_{b_{11}| h_i^{eff}}$ and $ P_{b_{12}| h_i^{eff}}$ are given in (\ref{b11}) and (\ref{b12}), respectively.


\subsection{Analysis of \texorpdfstring{$U_2$}{Lg}}
To calculate the average \gls{ber} of $U_2$, we need to calculate the probability of detection error of its two individual bits, $b_{21}$ and $b_{22}$, then taking the average of them. Figure \ref{16_4_sys}(c) shows the constellation mapping of $b_{21}$, where the blue points represent $b_{21} = 1$ while the red points represent having $b_{21} = 0$. The constellation mapping of $b_{22}$ is exactly the same as shown in \figref{16_4_sys}(c) but horizontally not vertically. Hence, $b_{21}$ and $b_{22}$ have similar detection error probabilities due to the symmetry of their constellation mappings, which makes the average \gls{ber} of $U_2$ equal to the probability of detection error of $b_{21}$.

From \figref{16_4_sys}(c), an error can occur when detecting $b_{21}$ if any of the blue constellation points moves to the decision regions of the red points or vice versa. To calculate the total average detection error probability of $b_{21}$, we derive the conditional error probability given that the received constellation point belongs to $C_i$, then we take the average over all the possible $8$ columns. However, due to the symmetry between the red columns and the blue columns in \figref{16_4_sys}(c), we only calculate the conditional error probability given $C_1$, $C_3$, $C_5$ and $C_7$, and then we take the average over them.

Assuming that the true received constellation point belongs to $C_1$, and given $h_i^{eff}$, the conditional error probability of $b_{21}$ can be expressed as
\begin{align} \label{c1}
& P_{b_{21}| h_i^{eff}, C_1}  =  \mathrm{Pr} \Big{(} \mathfrak{Re}(n) \in \{ [-\lambda_3+\delta_4, -\lambda_2+\delta_4]  \cup  \nonumber \\
& \quad [-\lambda_1+\delta_4, \delta_4]   \cup   [\lambda_1+\delta_4, \lambda_2+\delta_4]  \cup  [\lambda_3+\delta_4, \infty]  \}  \Big{)}  \nonumber  \\
& \quad  {=}  Q \left( \frac{-\lambda_3 {+} \delta_4}{\sigma_n} \right) - Q \left( \frac{-\lambda_2 {+} \delta_4}{\sigma_n} \right) +  Q \left( \frac{-\lambda_1 {+} \delta_4}{\sigma_n} \right)  \nonumber  \\
& \quad {-} Q \left( \frac{ \delta_4}{\sigma_n} \right) {+} Q \left( \frac{\lambda_1 {+} \delta_4}{\sigma_n} \right) {-}  Q \left( \frac{\lambda_2 {+} \delta_4}{\sigma_n} \right)  {+} Q \left( \frac{\lambda_3 {+} \delta_4}{\sigma_n} \right).
\end{align}
If the received constellation point belongs to $C_3$, the conditional error probability of detecting $b_{21}$ can be expressed as
\begin{align} \label{c3}
& P_{b_{21}| h_i^{eff}, C_3}  =  \mathrm{Pr} \Big{(} \mathfrak{Re}(n) \in \{ [-\lambda_3+\delta_2, -\lambda_2+\delta_2]  \cup  \nonumber \\
& \quad [-\lambda_1+\delta_2, \delta_2]   \cup   [\lambda_1+\delta_2, \lambda_2+\delta_2]  \cup  [\lambda_3+\delta_2, \infty]  \}  \Big{)}  \nonumber  \\
& \quad  {=}  Q \left( \frac{-\lambda_3 {+} \delta_2}{\sigma_n} \right) - Q \left( \frac{-\lambda_2 {+} \delta_2}{\sigma_n} \right) +  Q \left( \frac{-\lambda_1 {+} \delta_2}{\sigma_n} \right)  \nonumber  \\
& \quad {-} Q \left( \frac{ \delta_2}{\sigma_n} \right) {+} Q \left( \frac{\lambda_1 {+} \delta_2}{\sigma_n} \right) {-}  Q \left( \frac{\lambda_2 {+} \delta_2}{\sigma_n} \right)  {+} Q \left( \frac{\lambda_3 {+} \delta_2}{\sigma_n} \right).
\end{align}
In the case of receiving a constellation point that belongs to $C_5$, the conditional error probability of detecting $b_{21}$ can be given as
\begin{align} \label{c5}
& P_{b_{21}| h_i^{eff}, C_5}  =  \mathrm{Pr} \Big{(} \mathfrak{Re}(n) \in \{ [-\lambda_3-\delta_1, -\lambda_2-\delta_1]  \cup  \nonumber \\
& \quad [-\lambda_1-\delta_1, -\delta_1]   \cup   [\lambda_1-\delta_1, \lambda_2-\delta_1]  \cup  [\lambda_3-\delta_1, \infty]  \}  \Big{)}  \nonumber  \\
& \quad  {=}  Q \left( \frac{-\lambda_3 {-} \delta_1}{\sigma_n} \right) - Q \left( \frac{-\lambda_2 {-} \delta_1}{\sigma_n} \right) +  Q \left( \frac{-\lambda_1 {-} \delta_1}{\sigma_n} \right)  \nonumber  \\
& \quad {-} Q \left( \frac{ -\delta_1}{\sigma_n} \right) {+} Q \left( \frac{\lambda_1 {-} \delta_1}{\sigma_n} \right) {-}  Q \left( \frac{\lambda_2 {-} \delta_1}{\sigma_n} \right)  {+} Q \left( \frac{\lambda_3 {-} \delta_1}{\sigma_n} \right).
\end{align}
When the received constellation point belongs to $C_7$, the conditional error probability of detecting $b_{21}$ can be calculated as
\begin{align} \label{c7}
& P_{b_{21}| h_i^{eff}, C_7}  =  \mathrm{Pr} \Big{(} \mathfrak{Re}(n) \in \{ [-\lambda_3-\delta_3, -\lambda_2-\delta_3]  \cup  \nonumber \\
& \quad [-\lambda_1-\delta_3, -\delta_3]   \cup   [\lambda_1-\delta_3, \lambda_2-\delta_3]  \cup  [\lambda_3-\delta_3, \infty]  \}  \Big{)}  \nonumber  \\
& \quad  {=}  Q \left( \frac{-\lambda_3 {-} \delta_3}{\sigma_n} \right) - Q \left( \frac{-\lambda_2 {-} \delta_3}{\sigma_n} \right) +  Q \left( \frac{-\lambda_1 {-} \delta_3}{\sigma_n} \right)  \nonumber  \\
& \quad {-} Q \left( \frac{ -\delta_3}{\sigma_n} \right) {+} Q \left( \frac{\lambda_1 {-} \delta_3}{\sigma_n} \right) {-}  Q \left( \frac{\lambda_2 {-} \delta_3}{\sigma_n} \right)  {+} Q \left( \frac{\lambda_3 {-} \delta_3}{\sigma_n} \right).
\end{align}
Thus, the average \gls{ber} of $U_2$ given the effective channels, $h_i^{eff}$, can be calculated as
\begin{equation}  \label{ber_U2_2users}
P_{U_2| h_i^{eff}}  =  P_{b_{21}| h_i^{eff}} = \frac{1}{4} \sum_{m=0}^3  P_{b_{21}| h_i^{eff}, C_{2m + 1}},
\end{equation}
where $P_{b_{21}| h_i^{eff}, C_{2m + 1}}$ for $m=0, 1, 2$, and $3$ is given in (\ref{c1}), (\ref{c3}), (\ref{c5}), and (\ref{c7}), respectively.

After deriving the conditional BERs given the instantaneous effective channels of the users, the derivation of the average unconditional \gls{ber} expressions of $U_1$ and $U_2$ is presented in Sec. \ref{averaging_ch}. This is performed by averaging the expressions obtained in (\ref{ber_U1_2users}) and (\ref{ber_U2_2users}), respectively, over the \glspl{pdf} of the effective channels, $h_1^{eff}$ and $h_2^{eff}$.

\section{Design and BER analysis of the generalized \texorpdfstring{$K$}{Lg}-user \texorpdfstring{$M$}{Lg}-QAM uplink RIS-NOMA system} \label{general}
In this section, a framework is presented for deriving closed-form \gls{ber} expressions for the generalized uplink \gls{ris}-\gls{noma} system with an arbitrary number of users and arbitrary \gls{qam} modulation orders. Specifically, we extend our \gls{ber} analysis to the generalized system by specifically providing an algorithm that calculates the coefficients of the \gls{ber} expressions for an arbitrary number of users and \gls{qam} modulation orders. Then, we provide a methodology to derive the unconditional average \gls{ber} expressions of the generalized system by averaging over the \glspl{cf} of the effective channels of the $K$ users.

\subsection{Generalization for arbitrary number of users and QAM modulation orders}

In this subsection, we generalize our analysis by devising a methodology to obtain the \gls{ber} expressions for a general uplink \gls{ris}-\gls{noma} system with arbitrary number of users and arbitrary \gls{qam} modulation orders for the users. From Sec. \ref{2-user}, we have found that the conditional \gls{ber} expressions always take the form of a weighted sum of $Q(\cdot)$ functions having different linear combinations of the effective channels, $h_i^{eff}$, as arguments. Consequently, for a general $K$-user system, the conditional \gls{ber} expression of user $k$, given the effective channel values, $h_i^{eff}, \forall i$, can be expressed as
\begin{equation}  \label{general_ber}
BER_{U_k|h_1^{eff}, \dots, h_K^{eff}} = \sum_{q=1}^{N_k} c_{k,q} Q \left( \frac{ \sum_{i=1}^K a_{k,iq} h_i^{eff} }{\sigma_n} \right),
\end{equation}
where $N_k$ is the total number of $Q(\cdot)$ functions in the \gls{ber} expression of $U_k$, while the coefficients $c_{k,q}$ and $a_{k,iq}$ depend on the number of users, $K$, and the \gls{qam} modulation orders used by these users. Therefore, by devising an algorithm that can compute these coefficients for arbitrary number of users and \gls{qam} modulation orders, we can calculate the conditional \gls{ber} expressions for a generalized uplink \gls{ris}-\gls{noma} system.

Due to applying the proposed channel alignment scheme in Sec. \ref{ch_align}, the received superimposed uplink \gls{noma} constellation at the \gls{bs} is constructed as the constellation of downlink \gls{noma} systems. Therefore, utilizing the proposed channel alignment scheme allows us to use the methodology provided in Sec. III of \cite{hamad} for deriving the generalized \gls{ber} expressions of downlink \gls{noma} systems. Section III of \cite{hamad} provides a methodology to obtain the weights of the $Q(\cdot)$ functions, $c_{k,q}$, in (\ref{general_ber}), but does not calculate the coefficients, $a_{k,iq}$, which are inside the $Q(\cdot)$ functions. The reason for this is that, unlike uplink \gls{noma} systems, the constellation of downlink \gls{noma} systems experiences only one fading channel before reception at the user, where the signal received at the $k$th user can be given as
\begin{equation}
y_k = \left( \sum_{i=1}^K \sqrt{\rho_i} x_i \right) h_k + n_k,
\end{equation}
where $\rho_i$ is the power allocated to user $i$, while $h_k$ and $n_k$ are the fading channel and \gls{awgn} of the $k$th user, respectively. However, the methodology in \cite{hamad} calculates the error distances of the downlink \gls{noma} constellation which can be written as
\begin{equation}  \label{dist_equ}
\Delta_{k,q} = \sum_{i=1}^K a_{k,iq} \sqrt{\rho_i},
\end{equation}
where $a_{k,iq}, \forall i,q$ are the required coefficients for the $k$th user in (\ref{general_ber}). Hence, setting $\rho_i$ to some known values, we can compute the coefficients, $a_{k,iq}$, of the $k$th user by knowing the distances, $\Delta_{k,q}$, using \textbf{Algorithm \ref{coeff_alg}}. The $i$th element of the input vector, $\mathbf{b}$, of \textbf{Algorithm \ref{coeff_alg}} is the number of bits per symbol of the $i$th user, $b_i$, representing its modulation order, while $\boldsymbol{\rho}$ contains the power coefficients of the users, $\rho_i, \forall i$. The vector, $\mathbf{\Delta}_k$, contains all the error distances of the $k$th user, $\Delta_{k,q}, \forall q$. The output of \textbf{Algorithm \ref{coeff_alg}} are the matrices, $\mathbf{A}_1, \dots, \mathbf{A}_K$, which contain the required coefficients, $a_{k,iq}, \forall i,q$, that is, the element of the $q$th row and the $i$th column of $\mathbf{A}_k$ is $a_{k,iq}$.

\begin{algorithm}
\SetAlgoLined
\textbf{Inputs:} $\mathbf{b}$ and $\boldsymbol{\rho}$. \\
\For{k=1:K}{
Calculate $\mathbf{\Delta}_k^{(1)}$ using the method in \cite{hamad}, with inputs $\mathbf{b}$ and $\boldsymbol{\rho}$. \\
\For {$i = 1:K$}{
Set $\varepsilon$ to any small value, e.g., $\varepsilon = 0.1$; \\
Set $\boldsymbol{\rho}^\prime = \boldsymbol{\rho}$; \\
Set $\boldsymbol{\rho}^\prime(i) = \left( \sqrt{\boldsymbol{\rho}(i)} + \varepsilon \right)^2$; \\
Calculate $\mathbf{\Delta}_k^{(2)}$ using the method in \cite{hamad}, with inputs $\mathbf{b}$ and $\boldsymbol{\rho}^\prime$. \\
\For{$q = 1:\mathrm{length} \left( \mathbf{\Delta}_k^{(1)} \right)$}{
$\mathbf{A}_k(q,i) = \frac{ \mathbf{\Delta}_k^{(2)}(q) - \mathbf{\Delta}_k^{(1)}(q) } {\varepsilon}$;
}
}
}
\textbf{Return:} $\mathbf{A}_1, \mathbf{A}_2, \dots, \mathbf{A}_K$.
\caption{Calculation of $a_{k,iq}, \forall k,i,q$.}
\label{coeff_alg}
\end{algorithm}

The idea of \textbf{Algorithm \ref{coeff_alg}} is that the coefficient, $a_{iq}$, in (\ref{dist_equ}) can be calculated using the partial differentiation of $\Delta_{k,q}$ \gls{wrt} $\sqrt{\rho_i}$. Since $\Delta_{k,q}$ is a linear function in $\sqrt{\rho_i}$, then the partial derivative of $\Delta_{k,q}$ \gls{wrt} $\sqrt{\rho_i}$ can be calculated as
\begin{equation}
a_{k,iq} = \mathbf{A}_k(q,i) = \frac{\partial \Delta_{k,q}}{\partial \sqrt{\rho_i}} = \frac{\Delta_{k,q}^{(2)} - \Delta_{k,q}^{(1)}} {\sqrt{\rho_i}^{(2)} - \sqrt{\rho_i}^{(1)}}, \quad \forall k, i, q.
\end{equation}
The power coefficients, $\rho_i$, are set in \textbf{Algorithm \ref{coeff_alg}} as $\rho_i = 2^{\sum_{i+1}^K b_i}, \forall i < K$ and $\rho_K = 1$, because this setting prevents overlapping between the superimposed \gls{noma} constellation points.


\subsection{Averaging the BER expressions over the PDFs of the RIS effective channels}  \label{averaging_ch}
In this subsection, we provide a methodology to average the obtained conditional \gls{ber} expressions in (\ref{general_ber}) over the \glspl{pdf} of $h_i^{eff}, \forall i$. However, the random channel components, $\mathfrak{Re}(\gamma_{ij}), i \neq j$, are correlated since they contain common \gls{ris}-\gls{bs} channel coefficients, $\mathbf{g}_j$. For example, in the three-user scenario, the argument inside the $q$th $Q(\cdot)$ function of the \gls{ber} expression of the $k$th user can always be expressed as
\begin{align}  \label{x_equ}
X_{k,q} &= a_{k,1q} h_1^{eff} + a_{k,2q} h_2^{eff} + a_{k,3q} h_3^{eff}  \nonumber  \\
& \approx \widetilde{a}_{k,1q} \left( \sum\nolimits_{l=1}^{L_1} |h_{11,l}| |g_{1,l}| + \mathfrak{Re} \left(  \widetilde{\mathbf{g}}_{2}^T  \mathbf{h}_{12} + \widetilde{\mathbf{g}}_{3}^T  \mathbf{h}_{13} \right) \right)  \nonumber \\
& + \widetilde{a}_{k,2q} \left( \sum\nolimits_{l=1}^{L_2} |h_{22,l}| |g_{2,l}| + \mathfrak{Re} \left(  \widetilde{\mathbf{g}}_{1}^T  \mathbf{h}_{21} + \widetilde{\mathbf{g}}_{3}^T  \mathbf{h}_{23} \right) \right)  \nonumber \\
& + \widetilde{a}_{k,3q} \left( \sum\nolimits_{l=1}^{L_3} |h_{33,l}| |g_{3,l}| + \mathfrak{Re} \left(  \widetilde{\mathbf{g}}_{1}^T  \mathbf{h}_{31} + \widetilde{\mathbf{g}}_{2}^T  \mathbf{h}_{32} \right) \right),
\end{align}
where $\widetilde a_{k,iq} = a_{k,iq} \sqrt{\frac{P_i \eta_i}{\beta_i}}$, and $\widetilde{\mathbf{g}}_{i} = \mathbf{\Theta}_i \mathbf{g}_{i}$ where $\mathbf{\Theta}_i$ is the diagonal reflection coefficients matrix of the $i$th partition, $\mathcal{P}_i$. Now, we need to derive the \gls{cf} of the compound \gls{rv}, $X$, to use it to derive the average of $Q(X)$. However, the real components of (\ref{x_equ}) are correlated, since $\widetilde g_{1,l}$, $\widetilde g_{2,l}$ and $\widetilde g_{3,l}$ are common factors in them. Consequently, the terms of $X$ in (\ref{x_equ}) are rearranged in a way that puts them in the form of a sum of uncorrelated terms, as
\begin{align}  \label{x_arranged}
& X_{k,q} \approx  \sum_{i=1}^3  \widetilde{a}_{k,iq} \sum_{l=1}^{L_i} |h_{ii,l}| |g_{i,l}| 
\nonumber  \\
& \quad +  \mathfrak{Re} \Big ( \widetilde{\mathbf{g}}_1^T ( \widetilde{a}_{k,2q} \mathbf{h}_{21}  {+} \widetilde{a}_{k,3q} \mathbf{h}_{31} )  {+}  \widetilde{\mathbf{g}}_2^T ( \widetilde{a}_{k,1q} \mathbf{h}_{12} {+} \widetilde{a}_{k,3q} \mathbf{h}_{32} )  \nonumber \\
& \quad +  \widetilde{\mathbf{g}}_3^T ( \widetilde{a}_{k,1q} \mathbf{h}_{13} {+} \widetilde{a}_{k,2q} \mathbf{h}_{23} ) \Big).
\end{align}
Now, the \gls{cf} of $X_{k,q}$ can be easily derived by multiplying the \gls{cf} of the terms in (\ref{x_arranged}). Since the absolute value terms are summations of product of two Rayleigh \glspl{rv}, they can be modeled as Gamma \glspl{rv} having \glspl{cf} as in (\ref{gamma_approx}). For the real value terms, each of them is equivalent in distribution to
\begin{equation} \label{normal_sum}
S_i \triangleq \sqrt{\sum_{j=1, j \neq i}^3 |\widetilde a_{k,jq}|^2} \ \mathfrak{Re} \left( \widetilde {\mathbf{g}}_i^T \widehat{\mathbf{h}}_i \right),
\end{equation}
where both $\widetilde {\mathbf{g}}_i$ and $\widehat{\mathbf{h}}_i$ are \gls{iid} $\mathcal{CN}(\mathbf{0}, \mathbf{I})$. Since $S_i$ is a real sum of products of two independent $\mathcal{CN}(0,1)$, it can be modeled as a difference between two \gls{iid} Erlang distributions as in Proposition \ref{theo_1}. Finally, by induction for a general $K$-user system, the \gls{cf} of $X_{k,q}$ can be written using (\ref{gamma_approx}) and (\ref{diff_erlang}) as
\begin{equation}
\Phi_{X_{k,q}}\!(\!z\!) {=\!} \left( \prod_{i=1}^K \!\!\left( 1 {-} j^\prime a_{k,iq} \zeta_i z \right) ^{-N_i} \!\!\right)\!\!  \left( \prod_{i=1}^K \!\!\left( 1 {+} \frac{\kappa_i}{4} z^2 \right)^{-L_i}\!\!\right)\!\!,
\end{equation}
where $\zeta_i$ and $N_i$ are given in (\ref{zeta_and_N}), and $\kappa_i = \sum_{j=1, j \neq i}^K |\widetilde a_{k,jq}|^2$. Finally, Proposition \ref{theo_2} gives the final unconditional \gls{ber} of $U_k$.

\begin{proposition}  \label{theo_2}
By averaging (\ref{general_ber}) over the statistical distributions of $h_i^{eff}, \forall i$, the unconditional average \gls{ber} of $U_k$ can be given as
\[ BER_{U_k} {=} \sum_{q=1}^{N_k} \frac{c_{k,q}}{2} {+} \frac{c_{k,q}}{ \pi} \int_0^{\infty} \mathfrak{Re} \left( \frac{j^\prime e^{ \frac{-z^2}{2} } }{z}  \Phi_{X_{k,q}} \left( \frac{z}{\sigma_n} \right) \right) dz .\]
\end{proposition}

\renewcommand\qedsymbol{$\blacksquare$}

\begin{proof}
Please see Appendix \ref{appx_b}.
\end{proof}


\section{Average BER minimization based power allocation scheme} \label{opt}

In this section, we introduce an optimized uplink \gls{pa} scheme for the \gls{ris}-\gls{noma} system. The aim of this \gls{pa} is to eliminate the \gls{ber} floors that occur in \gls{sic}-based uplink \gls{noma} systems. Specifically, the transmit powers of the users are optimized to minimize the overall average \gls{ber} performance at the \gls{bs} for all users while adhering to individual uplink transmit power constraints. The optimization process is clearer and more robust when conducted in the log-log domain, where both the power values to be optimized, $P_k$, and the cost function are expressed in $\mathrm{dB}$. This transformation results in smoother numerical optimization and faster convergence when applying gradient descent-based optimization techniques to solve our problem. Therefore, the uplink \gls{pa} problem can be formulated as
\begin{subequations} \label{opt_prob}
\begin{align}
\min_{P_1, P_2, \dots, P_K} \quad & 10 \log_{10} \left( \sum_{k=1}^K BER_{U_k} \left( 10^{\frac{P_1}{10}}, \dots, 10^{\frac{P_K}{10}} \right) \right) \label{cost_fun} \\ 
{\mathrm {s.t.}} \quad & P_k  \leq  P_{\mathrm{dBm}}^{\mathrm{max}}, \qquad \forall k,
\end{align}
\end{subequations}
where $P_{\mathrm{dBm}}^{\mathrm{max}}$ represents the maximum available uplink transmit power in $\mathrm{dBm}$, and $BER_{U_k}$ in (\ref{cost_fun}) is the general \gls{ber} formula of $U_k$ given in Proposition \ref{theo_2}.

We use a gradient-based method to solve (\ref{opt_prob}) where the cost function along with the constraints are all continuous and have continuous first derivatives. To deal with the constraints, we shall transform (\ref{opt_prob}) into an unconstrained optimization problem by constructing the Lagrangian that is given as
\begin{equation}  \label{largrange}
L(\mathbf{p}, \xi_1, \dots, \xi_K) = f(\mathbf{p}) + \sum_{k=1}^K \xi_k (P_k - P_{dBm}^{\mathrm{max}}),
\end{equation}
where $\xi_k$ is the Lagrange multiplier that corresponds to the $k$th constraint, $f$ is the cost function in (\ref{opt_prob}), and $\mathbf{p} = [P_1, \dots, P_K]^T$. The Lagrangian in (\ref{largrange}) is optimized using Newton's gradient-based method, utilizing the Hessian operator of the Lagrangian. Given that the constraints are linear functions, the Hessian in this case only includes the Hessian of the original objective function $f$, as it represents the second derivative of $L$. Therefore, the Hessian of $L$ can be given as
\begin{equation}
\nabla^2 L = \nabla^2 f(\mathbf{p}),
\end{equation}
where $\nabla^2$ is the Laplacian operator or the second order differential operator. Newton's method is an iterative algorithm that is applied using a starting point $\mathbf{p}_0$ in an iterative manner until convergence to the optimum point. The update rule in each iteration can be given as
\begin{equation}
\mathbf{p}_{n+1} = \mathbf{p}_n - \left[ \nabla^2 f(\mathbf{p}_n) \right]^{-1} \nabla f(\mathbf{p}_n).
\end{equation}
The Lagrange multipliers, $\xi_k$, in (\ref{largrange}) are hyperparameters that are adjusted iteratively until convergence is reached. This involves repeatedly optimizing the Lagrangian, starting with initial values for $\xi_k$, which are updated in each iteration. Generally, we start with high values for $\xi_k$ and modify them based on how much the constraints are satisfied in every iteration. If $(P_k - P_{dBm}^{\mathrm{max}}) < 0$, the corresponding Lagrange multiplier, $\xi_k$, should be decreased. Conversely, if $(P_k-P_{dBm}^{\mathrm{max}}) > 0$, indicating a constraint violation, $\xi_k$ should be increased.

It is important to highlight that while the optimization process described above is iterative, it relies on the average channel statistics of the users rather than their instantaneous channel coefficients. This aspect simplifies the hardware implementation of the \gls{pa} algorithm. Specifically, the \gls{pa} problem outlined in (\ref{opt_prob}) is defined solely by the users' channel variances, $\sigma_k^2$, and the average received noise power, $\sigma_n^2$, all of which exhibit long coherence times. This method ensures that the proposed \gls{pa} strategy remains computationally efficient while providing minimized average \gls{ber} performance for all \gls{noma} users, making it suitable for real-world applications.


\section{Simulation Results} \label{Sim}

In this section, we provide simulation results for the achievable \gls{ber} of the introduced system under different operating conditions. We also validate our derived expressions using Monte-Carlo simulations with $10^7$ simulation runs. The \gls{ber} performance of the users is plotted against the transmit power while the Gaussian noise variance in the real or imaginary dimensions is set to unity, i.e., $\sigma_n^2 = 1$. It should be noted that the simulated BER is plotted up to $\mathrm{BER}=10^{-7}$ due to the limitation on the number of Monte-Carlo runs, whereas the analytical solution provides insights into the \gls{ber} for smaller than $10^{-7}$ values.

Moreover, the achievable \gls{ber} performance of the \gls{ris}-\gls{noma} system is compared to the \gls{ris}-\gls{oma} counterpart that applies \gls{tdma}. For fair comparison, the time slot allocated to a certain user in \gls{ris}-\gls{tdma}-\gls{oma} scenario, $T_{\mathrm{OMA}}$, equals the time slot in \gls{ris}-NOMA, $T_{\mathrm{NOMA}}$, divided by the total number of users, $K$, i.e., $T_{\mathrm{NOMA}}=K T_{\mathrm{OMA}}$. In addition, we fix the achievable data rate for all users by making the total number of bits transmitted by each user during $T_{\mathrm{OMA}}$ equal to the total number of bits transmitted by that user during $T_{\mathrm{NOMA}}$. By way of explanation, the modulation order for user $k$ in \gls{ris}-\gls{tdma}-\gls{oma} equals its modulation order in \gls{ris}-\gls{noma} scenario raised to power $K$, which guarantees that the total number of transmitted bits in both systems is equal for the same amount of consumed network resources. In addition, two scenarios of OMA, named OMA1 and OMA2, have been adopted to achieve a fair comparison. In OMA1, all partitions, that is, all reflectors in the RIS panel, are assigned to each user during its signaling period, whereas only one partition is assigned to a certain user during its signaling time in OMA2, which is similar to the assignment when NOMA signaling is applied. It should be mentioned that in NOMA signaling, the total number of channels to be estimated is $L=\sum_{i=1}^K L_i$, which is equal to the number of phase shifts to be compensated at the RIS, while in OMA1, the total number of channels to be estimated and the number of compensated phase shifts is $L \times K$. Hence, OMA1 adds considerable limitations because of the extreme computational complexity imposed due to the high number of channels to be estimated and compensated at the RIS, as well as the phase switching speed at the RIS to update the phase shifts. Therefore, OMA2 is also considered in our simulations as another benchmark, which is more fair and practical since it requires the same number of channel estimates and phase compensations as the NOMA scenario.

\begin{figure}
\centering
\includegraphics[width=\simwidth]{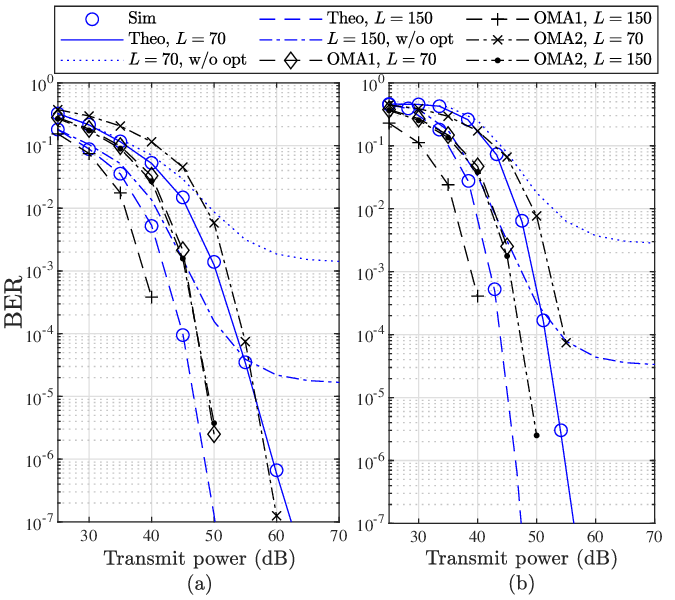}
\caption{BER performance comparison between a $2$-user RIS-NOMA system and the RIS-OMA counterpart under different $L$ values and $d_{U_1, I} = 20$, $d_{U_2, I} = 70$, $d_{I,B} = 30$: (a) $U_1$ ($16$-QAM), (b) $U_2$ ($4$-QAM).}
\vspace{-0.1in}
\label{fig1}
\end{figure}

Figure \ref{fig1} shows the \gls{ber} performance of a two-user uplink \gls{ris}-\gls{noma} system model against the transmission power of each user. The two users, $U_1$ and $U_2$, employ QAM with orders of $16$ and $4$ in \gls{noma} signaling, while they use modulation orders of $256$ and $16$, respectively, when \gls{tdma}-\gls{oma} is applied. Moreover, the users' distances to the \gls{ris} are $20$ m and $70$ m, respectively, while the \gls{ris}-\gls{bs} distance is $30$ m. The figure also compares the \gls{ris}-\gls{noma} with \gls{ris}-OMA-TDMA under two values for the number of \gls{ris} reflectors per user. In the first case, a number of $L_i=70$ reflectors is assigned to each user, whereas $L_i = 150$ reflectors are assigned to each user in the second case. As can be seen in the figure, without optimization (w/o opt), the \gls{ris}-\gls{noma} system suffers from a significant error floor which increases as $L$ decreases. For example, error floors in the BER of ${U}_1$ of approximately $1.5\times 10^{-3}$ and $1.5\times 10^{-5}$ are observed when $L_i$ is $70$ and $150$, respectively, while the error floors are about $3\times 10^{-3}$ and $3\times 10^{-5}$ for the second user. This error floor is attributed to the inter-user interference that dominates the \gls{awgn} at high \gls{snr} values. This error floor decreases when the number of \gls{ris} elements per user increases due to the \gls{sinr} enhancement gained by increasing $L_i$. However, as can be observed from the figure, the proposed \gls{pa} algorithm in Sec. \ref{opt} has managed to eliminate the error floor by controlling the transmit power of each user, which in turn controls the amount of interference imposed by each user on other users. By comparing the performance of the optimized \gls{noma} with \gls{tdma}-\gls{oma}1, it can be observed that the latter significantly outperforms \gls{noma} even though higher modulation orders are applied in \gls{tdma}-\gls{oma}. For instance, Fig. \ref{fig1}(a) shows that there is a power gain of almost $7$ dB in favor of \gls{tdma}-\gls{oma}1 at $\mathrm{BER}=10^{-5}$ and $L_i=70$, however, this gain decreases to about $2$ dB as $L_i$ increases to $150$. This superiority of TDMA-OMA comes at the expense of considerably higher complexity due to channel estimation overheads, since all partitions are assigned to every user in TDMA-OMA1. On the other hand, by comparing NOMA with TDMA-OMA2 which has similar channel estimation overheads, it can be observed in Fig. \ref{fig1} that, for $U_2$, NOMA outperforms TDMA-OMA2 for the whole range of SNR as shown in Fig. \ref{fig1}(b), and for a wide range of SNR values in Fig. \ref{fig1}(a) showing the BER of $U_1$. Finally, the figure shows a perfect match between the simulation results and the derived closed-form expressions, which confirms the accuracy of the analysis carried out in this paper.

\begin{figure}
\centering
\includegraphics[width=\simwidth]{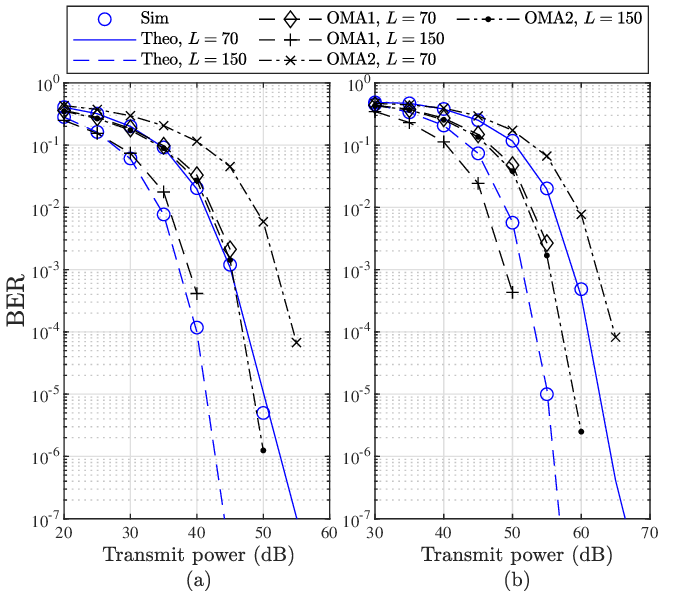}
\caption{BER performance comparison between a $2$-user RIS-NOMA system and the RIS-OMA counterpart under different $L$ values and $d_{U_1, I} = 20$, $d_{U_2, I} = 200$: (a) $U_1$ (16 QAM), (b) $U_2$ (4 QAM).}
\vspace{-0.1in}
\label{fig2}
\end{figure}

Figure \ref{fig2} shows the achievable \gls{ber} for the same simulation parameters as Fig. \ref{fig1} except from the distance of $U_2$ which has been increased to $200$ m in this figure. As can be seen in Fig. \ref{fig2}(a), increasing the distance of the second user, $U_2$, from the \gls{ris} reduces the interference on the first user, $U_1$, which also leads to a lower \gls{ber} for $U_1$. Therefore, as shown in Fig. \ref{fig2}(b), when \gls{noma} is applied, $U_1$ performs better than TDMA-OMA1 and TDMA-OMA2 for the whole SNR range. However, the performance of $U_2$ with \gls{tdma}-\gls{oma}1 is still better than that of \gls{noma}. Comparing Fig. \ref{fig1}(b) to Fig. \ref{fig2}(b), it can be observed that $U_2$ requires more power in the latter case to achieve the same BER in Fig. \ref{fig1}(b) due to the path loss imposed by long-distance transmission. Moreover, the results in Figs. \ref{fig1} and \ref{fig2} show that \gls{tdma}-\gls{oma}1 is generally preferred when low modulation orders are applied. However, it should be noted that in \gls{tdma}-\gls{oma}1, each user can make use of all elements of the \gls{ris}, which imposes additional channel estimation overheads, unlike \gls{noma} where the total number of reflectors is divided among users. Compared to TDMA-OMA2, which has the same channel estimation and phase compensation overheads as NOMA, the figure shows that NOMA is always preferable, as it provides lower BERs. The figure also shows a perfect match between the theoretical results and the simulation, which confirms the derivations carried out in this paper.

\begin{figure}
\centering
\includegraphics[width=\simwidth]{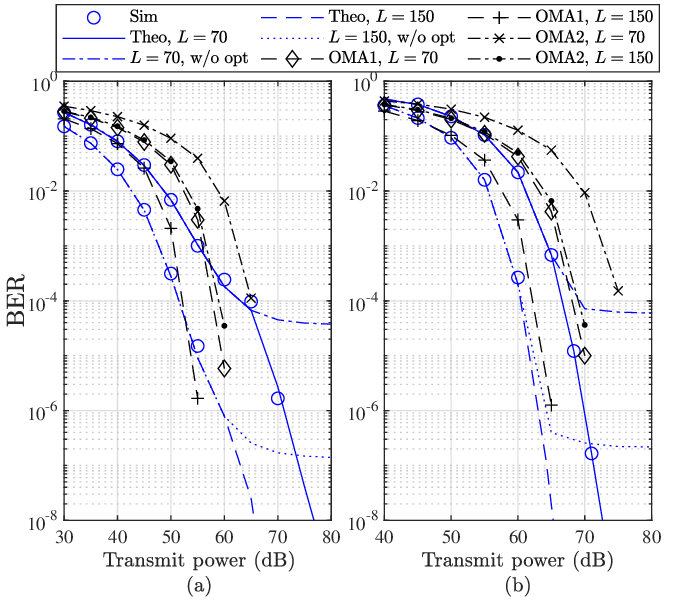}
\caption{BER performance comparison between a $2$-user RIS-NOMA system and the RIS-OMA counterpart under different $L$ values and $d_{U_1, I} = 20$, $d_{U_2, I} = 200$: (a) $U_1$ (64 QAM), (b) $U_2$ (16 QAM).}
\vspace{-0.1in}
\label{fig3}
\end{figure}

Figure \ref{fig3} shows the achievable BER for the same simulation parameters as Fig. \ref{fig2}, except for the modulation orders which have been increased to $64$ and $16$ for $U_1$ and $U_2$, respectively, for the \gls{noma} scenario, and $2^{12}$ and $2^8$ for $U_1$ and $U_2$, respectively, in the \gls{tdma}-\gls{oma} scenario. As can be observed in the figure, similar to Fig. \ref{fig1}, optimized resources manage to remove the BER floor. In addition, it can be seen from the figure that as the modulation orders increase, the performance of the optimized \gls{noma} becomes better than \gls{tdma}-\gls{oma}1 for the entire range of the transmit power of $U_2$, and for the low and mid ranges of the transmit power of $U_1$. For example, as can be seen from Fig. \ref{fig3}(a), the BER of $U_1$ using \gls{tdma}-\gls{oma} intersects with the BER of \gls{noma} at transmit powers of $53$ dB and $56$ dB for $L_i=150$ and $L_i=80$, respectively, where \gls{tdma}-\gls{oma}1 performs better for a higher transmit power regime. On the other hand, it can be clearly seen from the figure that optimized NOMA manages to provide lower BER than TDMA-NOMA2 for both users and for the whole range of SNR. The figure also confirms the analytical expressions derived in this paper for the BER.

\begin{figure*}
\centering
\includegraphics[width=2\columnwidth]{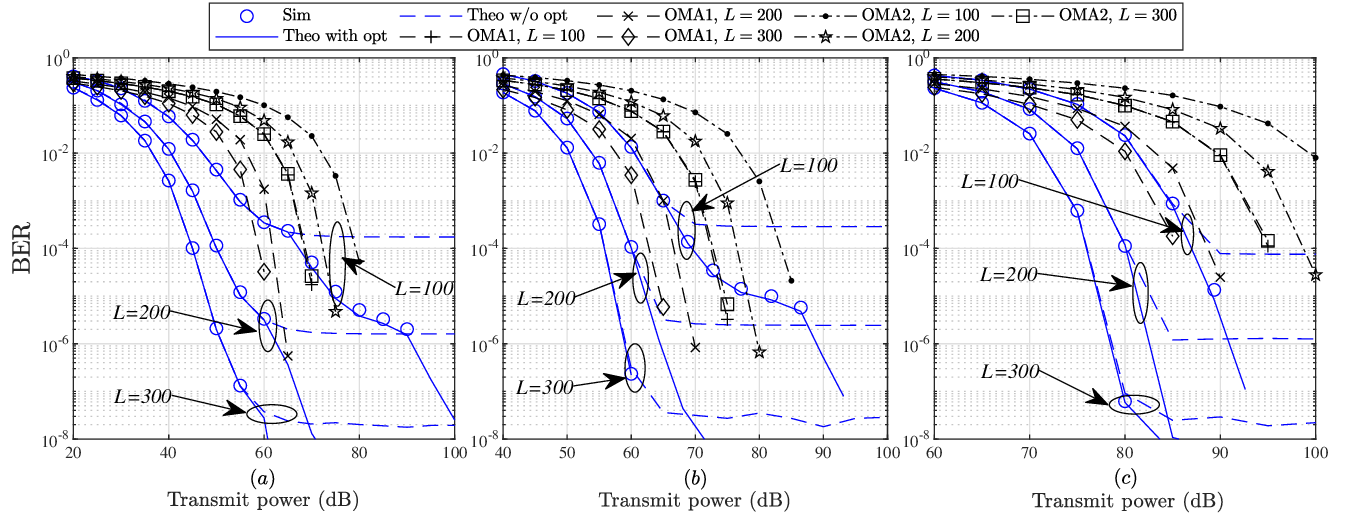}
\caption{BER performance comparison between a $3$-user RIS-NOMA system and the RIS-OMA counterpart under different $L$ values and $d_{U_1, I} = 20$, $d_{U_2, I} = 200$, $d_{U_3, I} = 2000$: (a) $U_1$ (64 QAM), (b) $U_2$ (16 QAM), (c) $U_3$ (16 QAM).}
\vspace{-0.1in}
\label{3-users}
\end{figure*}

To verify the analysis of the generalized case with multiple users, in Fig. \ref{3-users}, we present the achievable BER performance of a three-user uplink \gls{ris}-\gls{noma} scenario. The simulation parameters for $U_1$ and $U_2$ remain the same as in Fig. \ref{fig3}, except for the number of reflectors considered in the simulation setup where $L=\{100,200,300\}$ have been assumed for \gls{noma}. It should be noted that since each user uses the full potential of \gls{ris} in its time slot in the case of \gls{tdma}-\gls{oma}1, the equivalent number of reflectors in this case is $L=\{300,600,900\}$. The newly added $U_3$ employs $16$-QAM and is located at $2000$ m from the \gls{ris} location. As can be observed from the figure, there is a perfect match between theoretical results and simulation for the three-user scenario, which confirms the derivations carried out in this paper. The figure also shows that increasing the number of reflectors significantly reduces the error floor of all users, while the error floor can be mitigated by using the \gls{pa} algorithm introduced in this paper. Moreover, increasing the number of reflectors pushes the intersection points between the BER of \gls{noma} and \gls{tdma}-\gls{oma}1 to be at much higher values of the transmit power. This in turn gives superiority to the optimized \gls{noma} system over \gls{tdma}-\gls{oma}1 for a wide range of operating transmit power. For example, the BER of $U_3$ achieved using the optimized \gls{noma} is superior for the whole range of transmit power regardless $L$, and the superiority of the optimized \gls{noma} over the considered range of the transmit power for $U_1$ and $U_2$ is achieved when $L=\{200,300\}$. Moreover, when $L=100$, this superiority is achieved for transmit power less than $~68$ dB for $U_1$ and for transmit power less than $73$ dB for $U_2$. Furthermore, comparing RIS-NOMA with the TDMA-OMA2 counterpart, it can be clearly observed that RIS-NOMA has superior performance for all users and for the entire operating SNR range.



\section{Conclusions}  \label{Conc}
This paper investigated the achievable performance of uplink \gls{ris}-\gls{noma} system, where users transmit their data to the \gls{bs} using the same resource block and the \gls{bs} employs \gls{sic} to detect the data symbols of each user. Closed-form expressions for the achievable BER were derived to provide useful insights about the performance limits of \gls{ris}-\gls{noma}. The BER was compared to two \gls{tdma}-\gls{oma} counterparts to provide a comprehensive and fair comparison, where the first utilizes all partitions for each user at the expense of extra complexity, whereas the other assigns a partition to each user. Furthermore, a \gls{pa} algorithm was introduced to mitigate the error floor imposed by the inter-user interference of uplink \gls{noma}.

Theoretical results corroborated by simulations revealed that RIS-NOMA is preferred over TDMA-OMA2 for almost all the considered \gls{snr} regimes. On the other hand, the benefits gained by \gls{noma} could be limited compared to \gls{tdma}-\gls{oma}1 for low modulation orders and low numbers of reflectors, making \gls{tdma}-\gls{oma}1 able to outperform \gls{noma}. The main reason for this is attributed to the fact that in \gls{tdma}-\gls{oma}1, each user can use the full potential of the \gls{ris} panel, i.e., each user is assigned all reflectors, while the reflectors are divided among users in \gls{noma}. However, comparing \gls{noma} to TDMA-OMA1 is not fair since the latter imposes a much larger channel estimation overhead on the system than the former. Thus, the gain of TDMA-OMA1 comes at the expense of considerable computational and hardware complexity imposed by channel estimation and RIS phase compensation for large numbers of reflectors. Moreover, in practice, there is a practical constraint on the phase-switching speed of the reflectors, which makes it impossible to allocate the same reflector to a different user in subsequent signaling times. Furthermore, it was shown that by increasing the number of reflectors and/or increasing the modulation order, \gls{noma} can have superior performance compared to both TDMA-OMA scenarios for a wide range, and in some cases the whole range, of the transmit power.


\appendices

\section{Proof of Proposition 1}  \label{appx_a}

In this appendix, the \gls{cf} of $\mathfrak{Re}(\widehat{\gamma}_{ij})$, where $\widehat{\gamma}_{ij}$ is given in (\ref{normalized_rand_comp}), is derived. Assuming that $\widetilde{g}_{j,l} = e^{j^\prime \theta_{j,l}} g_{j,l}$, we can say that $\widetilde{g}_{j,l}$ is still $\mathcal{CN}(0,1)$. The reason for this fact is that $\arg(\widetilde{g}_{j,l}) \sim U(0, 2\pi)$ and it is independent of $\arg(g_{j,l})$, since both $\theta_{j,l}$ and $\arg(g_{j,l})$ are independent $U(0, 2\pi)$ \glspl{rv}. Therefore, $\widehat{\gamma}_{ij}$ is the sum of the products of \gls{iid} pairs of complex normal \glspl{rv} with zero mean and unit variance. To derive the statistical distribution of the real and imaginary parts of $\widehat{\gamma}_{ij}$, let us express each product term in the sum as
\begin{align}
T_l & = h_{ij,l} \widetilde{g}_{j,l} = (a_{ij,l} + j^\prime b_{ij,l}) (c_{j,l} + j^\prime d_{j,l})  \nonumber \\
& = (a_{ij,l} c_{j,l} - b_{ij,l} d_{j,l}) + j^\prime (b_{ij,l} c_{j,l} + a_{ij,l} d_{j,l}),
\end{align}
where $a_{ij,l}$ and $b_{ij,l}$ are the real and imaginary components of $h_{ij,l}$, respectively, while $c_{j,l}$ and $d_{j,l}$ are the real and imaginary components of $\widetilde{g}_{j,l}$, respectively. It should be noted that $a_{ij,l}$, $b_{ij,l}$, $c_{j,l}$ and $d_{j,l}$ are \gls{iid} Gaussian \glspl{rv} with zero mean and variance of $0.5$. Each product of two Gaussian \glspl{rv} can be written as
\begin{equation}
a_{ij,l} c_{j,l} = \frac{1}{4} \left[ (a_{ij,l} + c_{j,l})^2 - (a_{ij,l} - c_{j,l})^2 \right].
\end{equation}
Assuming that $G_1 = a_{ij,l} + c_{j,l}$ and $G_2 = a_{ij,l} - c_{j,l}$, the random vector $[G_1, G_2]^T$ can be written as 
\begin{equation}
\begin{bmatrix} G_1\\ G_2 \end{bmatrix}  = \begin{bmatrix} 1 & 1 \\ 1 & -1 \end{bmatrix} \begin{bmatrix} a_{ij,l} \\ c_{j,l} \end{bmatrix} = \mathbf{A} \begin{bmatrix} a_{ij,l} \\ c_{j,l} \end{bmatrix},
\end{equation}
where $\mathbf{A} = \begin{bmatrix} 1 & 1 \\ 1 & -1 \end{bmatrix}$. The random vector $[G_1, G_2]^T$ is a Gaussian random vector, since it is a linear transformation of another Gaussian random vector. The covariance matrix of $[a_{ij,l}, c_{j,l}]^T$ is $0.5 \mathbf{I}_2$, where $\mathbf{I}_2$ is the identity matrix of size $2$. Hence, the covariance matrix of $[G_1, G_2]^T$ can be given as \cite{papoulis}
\begin{equation}
\mathrm{cov}([G_1, G_2]^T) = 0.5 \mathbf{A} \mathbf{I}_2 \mathbf{A}^T = \mathbf{I}_2.
\end{equation}
Therefore, $G_1$ and $G_2$ are uncorrelated \glspl{rv}, since the covariance matrix of $[G_1, G_2]^T$ is diagonal. Consequently, $G_1$ and $G_2$ are independent since they are jointly Gaussian and uncorrelated \glspl{rv}.

Similarly, assuming that $F_1 = b_{ij,l} + d_{j,l}$ and $F_2 = b_{ij,l} - d_{j,l}$, the product $b_{ij,l} d_{j,l}$ can be written as
\begin{equation}
b_{ij,l} d_{j,l} = \frac{1}4{} \left( F_{1,l}^2 - F_{2,l}^2 \right),
\end{equation}
where both $F_1$ and $F_2$ are independent Gaussian \glspl{rv}. Hence, the real value of $\widehat{\gamma}_{ij}$ can be expressed as
\begin{equation} \label{erlang_diff}
\mathfrak{Re}(\widehat{\gamma}_{ij}) {=} \sum_{l=1}^{L_j} \mathfrak{Re}(T_l) {=} \frac{1}{4} \sum_{l=1}^{L_j} \left[ \left( G_{1,l}^2 + F_{2,l}^2 \right) {-} \left( G_{2,l}^2 + F_{1,l}^2 \right) \right].
\end{equation}
The $l$th term $G_{1,l}^2 {+} F_{2,l}^2$ in (\ref{erlang_diff}) can be modeled as an exponential \gls{rv} since it is the sum of two squared \gls{iid} Gaussian \glspl{rv}. The expected value of such exponential \gls{rv} can be calculated as
\begin{align}
\mu_{\mathrm{exp}} &= \mathbb{E}(G_{1,l}^2 + F_{2,l}^2) = \mathrm{var}(G_{1,l}) + \mathrm{var}(F_{2,l})  \nonumber \\
&= \mathrm{var}(a_{ij,l} + c_{j,l}) + \mathrm{var}(b_{ij,l} - d_{j,l}) = 2.
\end{align}
Similarly, $G_{2,l}^2 {+} F_{1,l}^2$ is an exponential \gls{rv} with a mean value of $2$. Since an $L_j$ sum of \gls{iid} exponential \glspl{rv} is modeled as an Erlang \gls{rv} \cite{papoulis}, $\mathfrak{Re}(\widehat{\gamma}_{ij})$ in (\ref{erlang_diff}) can be modeled as a difference between two \gls{iid} Erlang \glspl{rv} both having a shape parameter of $L_j$ and a scale parameter of $0.5$. Therefore, by using the \gls{cf} of an Erlang \gls{rv} \cite{papoulis}, the \gls{cf} of $\mathfrak{Re}(\widehat{\gamma}_{ij})$ can be given as in Proposition \ref{theo_1}.


\section{Proof of Proposition 2}  \label{appx_b}

Let $X$ be a \gls{rv} and $g(x)$ be a real one-dimensional function. Then, the expected value, $\mathbb{E}[g(X)]$, can be calculated using the formula \cite{annamalai}
\begin{equation}
\mathbb{E}[g(x)] = \frac{1}{\pi} \int_{0}^{\infty} \mathfrak{Re} \left\{ G(z) \Phi(z) \right\} dz,
\end{equation}
where $G(z)$ is the \gls{ft} of the function $g(x)$ and $\Phi(x)_X$ is the \gls{cf} of the \gls{rv}, $X$. Therefore, applying the above rule to average the conditional \gls{ber} in (\ref{general_ber}), the average unconditional \gls{ber} of $U_k$ can be calculated as
\begin{equation}  \label{app2_eq}
BER_{U_k} = \sum_{q=1}^{N_k}  \frac{c_{k,q}}{ \pi} \int_0^{\infty} \mathfrak{Re} \left( \mathfrak{Q}(z)  \Phi_{X_{k,q}} \left( \frac{z}{\sigma_n} \right) \right) dz,
\end{equation}
where $\mathfrak{Q}(z)$ is the \gls{ft} of the $Q$-function, which can be written as
\begin{equation}
\mathfrak{Q}(z) = \mathcal{F}( Q(t) ) = \mathcal{F} \Big( \int_t^{\infty} e^{-y^2/2} dy \Big),
\end{equation}
where $\mathcal{F}(.)$ is the \gls{ft} operator. Using the integral property of the \gls{ft} in \cite{brian} and getting $\mathcal{F}(e^{-t^2/2})$ from the \gls{ft} table, $\mathfrak{Q}(z)$ can be given as
\begin{equation}
\mathfrak{Q}(z) = \pi \delta(z) + j^\prime \frac{e^{-z^2/2}}{z},
\end{equation}
where $\delta(.)$ is the Dirac delta function. Substituting $\mathfrak{Q}(z)$ into (\ref{app2_eq}), we can obtain the result of Proposition \ref{theo_2}.

\footnotesize

\bibliographystyle{IEEEtran}
\bibliography{MyLib}

\end{document}

%% file: abbrev.tex
\newacronym{noma}{NOMA}{non-orthogonal multiple access}
\newacronym{sic}{SIC}{successive interference cancellation}
\newacronym{jmld}{JMLD}{joint maximum likelihood decoding}
\newacronym{mrc}{MRC}{maximum ratio combining}
\newacronym{bpsk}{BPSK}{binary phase shift keying}
\newacronym{qam}{QAM}{quadrature amplitude modulation}
\newacronym{pam}{PAM}{pulse amplitude modulation}
\newacronym{psk}{PSK}{phase shift keying}
\newacronym{mpsk}{M-PSK}{M-phase shift keying}
\newacronym{siso}{SISO}{single input single output}
\newacronym{mimo}{MIMO}{multiple input multiple output}
\newacronym{miso}{MISO}{multiple input single output}
\newacronym{simo}{SIMO}{single input multiple output}
\newacronym{iid}{i.i.d.}{independent and identically distributed}
\newacronym{awgn}{AWGN}{additive white Gaussian noise}
\newacronym{los}{LoS}{line-of-sight}
\newacronym{nlos}{NLoS}{non-LoS}
\newacronym{sinr}{SINR}{signal-to-interference-plus-noise-ratio}
\newacronym{bs}{BS}{base station}
\newacronym{ee}{EE}{energy efficiency}
\newacronym{se}{SE}{spectral efficiency}
\newacronym{qos}{QoS}{quality-of-service}
\newacronym{csi}{CSI}{channel state information}
\newacronym{mui}{MUI}{multi-user interference}
\newacronym{ber}{BER}{bit error rate}
\newacronym{ser}{SER}{symbol error rate}
\newacronym{qpsk}{QPSK}{quadrature phase shift keying}
\newacronym{rv}{RV}{random variable}
\newacronym{pdf}{PDF}{probability density function}
\newacronym{pmf}{PMF}{probability mass function}
\newacronym{sep}{SEP}{symbol error probability}
\newacronym{ue}{UE}{user equipment}
\newacronym{pep}{PEP}{pairwise error probability}
\newacronym{sdr}{SDR}{software defined radio}
\newacronym{ris}{RIS}{reconfigurable intelligent surface}
\newacronym{clt}{CLT}{central limit theorem}
\newacronym{cf}{CF}{characteristic function}
\newacronym{snr}{SNR}{signal-to-noise-ratio}
\newacronym{uav}{UAV}{unmanned ariel vehicles}
\newacronym{star-ris}{STAR-RIS}{simultaneously transmitting and reflecting reconfigurable intelligent surface}
\newacronym{tdma}{TDMA}{time division multiple access}
\newacronym{oma}{OMA}{orthogonal multiple access}
\newacronym{pa}{PA}{power allocation}
\newacronym{wrt}{w.r.t.}{with respect to}
\newacronym{iot}{IoT}{Internet of Things}
\newacronym{ft}{FT}{Fourier transform}